\documentclass{aa}  

\usepackage{graphicx}
\usepackage{txfonts}
\usepackage{longtable}
\usepackage{mhchem}
\usepackage{ulem}
\usepackage{color}
\usepackage[colorlinks = true, linkcolor = blue, anchorcolor = red, citecolor = blue, filecolor = blue, urlcolor = blue]{hyperref}

\begin{document} 

   \title{Production of linear alkanes via the solid-state hydrogenation of interstellar polyynes}
   
  \titlerunning{Production of linear alkanes by solid-state hydrogenation of interstellar polyynes}
  
   \author{G. Fedoseev\inst{1,2}, X. Li\inst{1,2}, G. A. Baratta\inst{3}, M. E. Palumbo\inst{3}, \and K.-J. Chuang\inst{4}}
 
   \institute{Xinjiang Astronomical Observatory, Chinese Academy of Sciences, Urumqi 830011, China\\
     \email{gleb@xao.ac.cn}
    \and
    Xinjiang Key Laboratory of Radio Astrophysics, Urumqi 830011, China
    \and
    INAF – Osservatorio Astrofisico di Catania, via Santa Sofia 78, 95123 Catania, Italy
    \and
    Laboratory for Astrophysics, Leiden Observatory, Leiden University, P.O. Box 9513, NL-2300 RA Leiden, the Netherlands}
    \authorrunning{G. Fedoseev et al.}  
   \date{Received XX XX, XX; accepted XX XX, XX}

% \abstract{}{}{}{}{} 
% 5 {} token are mandatory
 
  \abstract
   {Highly unsaturated carbon chains, including polyynes, have been detected in many astronomical regions and planetary systems. With the success of the QUIJOTE survey of the Taurus Molecular Cloud-1 (TMC-1), the community has seen a "boom" in the number of detected carbon chains. On the other hand, the Rosetta mission revealed the release of fully saturated hydrocarbons, C$_3$H$_8$, C$_4$H$_{10}$, C$_5$H$_{12}$, and (under specific conditions) C$_6$H$_{14}$ with C$_7$H$_{16}$, from the comet 67P/Churyumov-Gerasimenko. The detection of the latter two is attributed to dust-rich events. Similarly, the analysis of samples returned from asteroid Ryugu by Hayabusa2 mission indicates the presence of long saturated aliphatic chains in Ryugu's organic matter. }
   {The surface chemistry of unsaturated carbon chains under conditions resembling those of molecular clouds can provide a possible link among these independent observations. However, laboratory-based investigations to validate such a chemistry is still lacking. In the present study, we aim to experimentally verify the formation of fully saturated hydrocarbons by the surface hydrogenation of C$_{2n}$H$_2$ ($n>1$) polyynes under ultra-high vacuum conditions at 10 K. }
   {We undertook a two-step experimental technique. First, a thin layer of C$_2$H$_2$ ice was irradiated by UV-photons ($\geq$ 121 nm) to achieve a partial conversion of C$_2$H$_2$ into larger polyynes: C$_4$H$_2$ and C$_6$H$_2$. Afterwards, the obtained photoprocessed ice was exposed to H atoms to verify the formation of various saturated hydrocarbons.}
   {In addition to C$_2$H$_6$, which was investigated previously, the formation of larger alkanes, including C$_4$H$_{10}$ and (tentatively) C$_6$H$_{14}$, is confirmed by our study. A qualitative analysis of the obtained kinetic data indicates that hydrogenation of HCCH and HCCCCH triple bonds proceeds at comparable rates, given a surface temperature of 10 K. This can occur on the timescales typical for the dark cloud stage. A general pathway resulting in formation of other various aliphatic organic compounds by surface hydrogenation of N- and O-bearing polyynes is also proposed. We also discuss the astrobiological implications and the possibility of identifying alkanes with JWST.}
   {}

               
   \maketitle
%
%________________________________________________________________

\section{Introduction}
Interstellar polyynes are a large class of chemical compounds that can be described with the common formula R$_1$–(C)$_{2n}$–R$_2$, where $n>1$. Due to the presence of the conjugated unsaturated triple bonds, these chemical compounds are typically linear in shape and, along with the simplest alkynes HCCH (C$_2$H$_2$, acetylene) and HCCCH$_3$ (C$_3$H$_4$ propyne), they are often referred as "carbon chains." Polyynes and other carbon chain molecules comprise the large fraction of species identified to date in the inter- and circumstellar medium (ISM and CSM). \cite{Taniguchi2024Ap&SS} reported the detection of more than 100 carbon-chain species. This accounts for about a third of species currently identified in the ISM. 

Polyynes and carbon-chain molecules are ubiquitous in the ISM. They have been detected in extragalactic sources, photodissociation regions (PDRs), planetary nebulae, and the envelopes around carbon-rich AGB stars \citep{Pardo2005ApJ, Pardo2022A&A, Cuadrado2015A&A, Martín2021A&A, Taniguchi2024Ap&SS}. Observations of the Taurus Molecular Cloud-1 (TMC-1) using the 40-meter Yebes telescope and the 100-meter Green Bank telescope resulted in the identification of many new gas phase carbon chain constituents. These developments have raised questions about their associated formation and destruction routes \citep{Loomis2021NatAs, Cernicharo2021A&Ab, Cernicharo2021A&Aa, Cernicharo2022A&A, Fuentetaja2022A&Ab, Fuentetaja2022A&Aa, Taniguchi2016ApJ, Taniguchi2017ApJ, Taniguchi2019ApJ}. A steady growth in the number of carbon chain molecules identified towards protostars and protoplanetary disks can be outlined \citep{Sakai2013ChRv, Bergner2018ApJ, Ilee2021ApJS, Guzman2021ApJS, vanDishoeck2023FaDi}.

It is generally assumed that carbon chain molecules in star-forming regions are primarily produced in the gas phase through various ion-molecule and neutral-neutral reaction sequences. Astrochemical numerical simulations have a clear focus on the introduction of new gas-phase formation routes explaining the abundance of the newly observed species. Sometimes, the introduction of new gas-phase consumption routes is required to avoid the overproduction of one of the observed intermediate species. Even though solid-state reactions on the surface of interstellar dust grains are often required to produce the initial precursors for these carbon chain species gas-phase formation routes, systematic studies of carbon chain reactivity on the grain surface are lacking.

The dust temperature in the densest regions of molecular dark clouds drops to as low as $\sim$10 K, resulting in the continuous freeze-out of the species, including atoms and molecules previously formed in the gas phase. The most prominent example is the freeze out of CO molecules produced in a sequence of gas-phase reactions starting from C$^+$ ions. A CO freeze-out is the initial point for the formation of various complex organic molecules (COMs) as well as carbon chain oxides \citep{Gerakines2001Icar, Öberg2009A&A, Modica2010A&A, Maity2015PCCP, Fedoseev2015MNRAS, Fedoseev2022ApJ, Chuang2017MNRAS, Qasim2019A&Aa, Urso2019A&A}. Surface interactions of accreting CO molecules and H-atoms result in the formation of CH$_3$OH ice and organic molecules as complex as glycerol, a key component of lipids and the membranes of living cells \citep{Fedoseev2017ApJ}. Unsaturated triple bonds comprising polyynes have high reactivities with accreting H atoms and other open-shell species, such as free radicals, even at 10 K temperature \citep{Molpeceres2022A&A}. Indeed, dedicated laboratory studies performed by \cite{Hiraoka2000ApJ} and \cite{Kobayashi2017ApJ} revealed high reactivity of the simplest alkyne C$_2$H$_2$ (HCCH, acetylene) with H atoms, resulting in the formation of C$_2$H$_4$ (CH$_2$CH$_2$, ethylene) and C$_2$H$_6$ (CH$_3$CH$_3$, ethane). Subsequently, \cite{Qasim2019ESCb} reported similar results for the surface hydrogenation of C$_3$H$_4$ (HCCCH$_3$, propyne). The formation of C$_3$H$_6$ (CH$_2$CHCH$_3$, propene) followed by the ultimate formation of the fully saturated C$_3$H$_8$ (CH$_3$CH$_2$CH$_3$, propane) at 10 K. Saturation of triple bonds is also observed in the presence of other radicals upon (non)energetic processing. \cite{Chuang2020A&A, Chuang2021A&A} and \cite{Qasim2019ESCb} investigated the solid-state reactivity of C$_2$H$_2$ and C$_3$H$_4$ with OH (hydroxyl) radicals, a well-known intermediate in forming H$_2$O ice. It was found that the addition of OH radicals to the triple CC bonds can proceed at 10 K. The activation barrier for this reaction is expected to be lower than that for the similar addition of H atoms. Moreover, the addition of OH radicals to the double CC bonds proceeds with even lower activation barriers, see \cite{Qasim2019ESCb} and \cite{Molpeceres2022A&A} for more details. The final products obtained after the simultaneous exposure of C$_2$H$_2$ and C$_3$H$_4$ to OH radicals and H atoms are various alcohols and aldehydes (ketones). These new reaction routes have been used to explain recent detections of i-C$_3$H$_7$OH (2-propanol) in Sgr B2(N2) in \cite{Belloche2022A&A}. They used a modified rate equation code accounting for the non-diffusive reactions, such as the interactions of species with immobile OH radicals. Similarly the routes provided by \cite{Qasim2019A&Aa} have been suggested to explain the first detection of n-C$_3$H$_7$OH (n-propanol; \citealt{Jimenez-Serra2022A&A}). So far, the simplest alkynes HCCH and HCCCH$_3$ are the only representatives of H-(C)$_{2n}$H and H-(C)$_{2n}$-CH$_3$ rows of alkynes and polyynes, whose surface reactivity with H atoms and OH radicals are systematically investigated.

This work aims to provide the first experimental confirmations for surface hydrogenation of H-(C)$_{2n}$H ($n>1$) polyynes under the conditions relevant to the early stages of star formation. Based on the outcome of the aforementioned experimental and theoretical studies, we expect that hydrogenation of H-(C)$_{2n}$H ($n>1$) polyynes results in the intermediate formation of various semi-saturated and fully saturated linear alkanes described by the formula C$_{2n}$H$_{4n+2}$. This will result in the presence of linear alkanes on the surface of icy grains already at prestellar stages. The partial survival of the pristine ice through the later stages of star formation would enrich celestial bodies with alkanes. Linear alkanes, including C$_2$H$_6$, C$_4$H$_{10}$, and C$_6$H$_{14}$, have been identified in comet 67P/Churyumov–Gerasimenko with high abundances of $\sim$5 $\times$ 10$^{-1}$ (C$_2$H$_6$) and $\sim$5 $\times$ 10$^{-3}$ (C$_4$H$_{10}$) with respect to the H$_2$O \citep{Schuhmann2019A&A, Altwegg2017MNRAS}. Moreover, the Hayabusa2 mission reported the presence of the saturated aliphatic chains identified in Ryugu’s organic matter \citep{Yabuta2023Sci}.

The surface chemistry of H-(C)$_{2n}$H polyynes with $n>1$ has not been  investigated as systematically in the laboratory due to the difficulties associated with the acquisition and preservation of polyyne samples. Experimental samples synthesized in the laboratory are kept at liquid-N$_2$ temperature to avoid spontaneous explosive decomposition (see \cite{Khlifi1995JMoSp} and \cite{Kim2009ApJS}, for example). In the present study, a two-step approach has been utilized. For the first time, a mixture of polyynes was obtained in situ on the substrate and the reactivity of this mixed polyyne ice with H-atoms was investigated in the same experimental setup. 

Section \ref{EXPERIMENTAL} describes the applied experimental methodology and utilised ultra-high vacuum (UHV) cryogenic setup. In Sect. \ref{RESULTS AND DISCUSSIONS}, the experimental results are presented and discussed. Section \ref{ASTROCHEMICAL IMPLICATIONS AND CONCLUSIONS} focuses on the astrochemical implications and offers our conclusions.

%__________________________________________________________________

\section{Experimental}\label{EXPERIMENTAL}
All experiments are performed using an ultra-high vacuum (UHV) setup, SURFRESIDE$^{\rm UV}$, with a UV-photon source module. The original SURFRESIDE design is optimized to study various astrochemically relevant solid-state atom- and radical-addition reactions at temperatures as low as 10 K, as described in \cite{Ioppolo2013RScI}. The implementation of a UV-photon source allows us to make a quantitative comparison  between the outcome of these "non-energetic" reaction routes, i.e., the reaction routes between the species in the thermal equilibrium with the surface, and the outcome of UV photon-induced "energetic" reaction routes \citep{Chuang2017MNRAS}. The description of the upgrade of the setup is presented in \cite{Fedoseev2016MNRAS} and \cite{Chuang2018A&Aa}. In this work, a UV-photon source is used to produce the thermodynamically unstable species directly in the ice samples. This allows for the study of the chemical reactivity between species that cannot be introduced into UHV setup by other means \cite[see, e.g.,][]{Butscher2015MNRAS, Butscher2016A&A, Butscher2017PCCP}. Following this latter approach, a mixture of unsaturated hydrocarbons, including unstable C$_4$H$_2$ and C$_6$H$_2$, can be synthesized in situ by UV irradiation of C$_2$H$_2$ ice. Then the reactivity of the newly produced hydrocarbons with H atoms is investigated following the exposure of the ice samples to an H-atom beam. It should be noted that this experimental approach does not aim to mimic the exact composition of ice mantles in the ISM or their complex interactions with impinging UV-photons and H-atoms. Instead, it is aimed at an experimental verification that the hydrogenation of polyynes occurs on a 10 K surface by accreting H atoms using astronomically relevant H-atom fluences.

\subsection{Experimental setup}
The SURFRESIDE$^{\rm UV}$ setup is composed of a UHV main chamber with a typical base pressure of $\sim$10$^{-10}$ mbar. The background residual gas mainly comprises H$_2$, while the accretion rate of background H$_2$O is negligible ($<$6.3 $\times$ 10$^{10}$ molecules cm$^{-2}$ s$^{-1}$, \cite{Chuang2018ApJb} for the period of the experiment. A rotatable gold-plated copper substrate is mounted on the cold tip of a helium closed-cycle cryostat that is positioned in the center of the UHV main chamber. The temperature of the substrate is regulated in the range between 8 and 450 K by means of resistive heating and monitored using two silicone diodes with the absolute accuracy of $\pm$ 0.5 K. The C$_2$H$_2$:He gas mixture or Ar gas is admitted to the main chamber through high-precision all-metal leak valves with the 22$^{\circ}$ angle to the substrate surface normal. Two independent dosing lines are used to keep these gases for the whole extent of the experiment. A mixture of C$_2$H$_2$ in He "bathing" gas is used to avoid the risk of spontaneous C$_2$H$_2$ decomposition. The substrate temperature is maintained far above the He condensation point. This prevents the formation of He ice in all experiments. The mixed beam of H atoms and undissociated H$_2$ molecules is obtained by thermal cracking of H$_2$ molecules, using a Hydrogen Atom Beam Source \citep[HABS;][]{Tschersich2000JAP}. HABS is mounted in an independent UHV chamber separated from the main chamber by a UHV shutter. A U-bent quartz pipe is placed along the path of H-atom beam to efficiently quench the high translational or ro-vibrational excited states of H-atoms and H$_2$-molecules prior to their impact on the substrate under the incident angle of 45$^{\circ}$. The application of two independent commercial H-atom beam sources allows for the crosscheck of the obtained results. UV-photons are generated by a microwave-discharged hydrogen flowing lamp (MDHL) attached to a MgF$_2$ viewport (with a cutoff at 115 nm) on the main chamber. The MgF$_2$ viewport is positioned in front of the substrate parallel to the substrate plane. This configuration allows incident photons to cover uniformly the whole area of the 2.5 $\times$ 2.5 cm$^2$ substrate. The H-atom and UV-photon fluxes are calibrated in situ, and the details of this procedure are available from \cite{Ioppolo2013RScI} as well as \cite{Fedoseev2016MNRAS} and \cite{Ligterink2015A&A}, respectively. The H-atom flux utilised in the present study amounts to 2.4 $\times$ 10$^{13}$ atoms cm$^{-2}$ s$^{-1}$. The lower limit on UV-photon flux at the substrate plane is estimated to be $\sim$1.3 $\times$ 10$^{13}$ cm$^{-2}$ s$^{-1}$. The example of UV emission spectra obtained for the used F-type MDHL is available in \cite{Chen2014ApJ} and \cite{Ligterink2015A&A}. The following commercially available reagents are used in the experiments: a 5\% mixture of HCCH in He (Praxair), H$_2$ (Linde, 5.0), and Ar (Linde, 5.0).

\begin{figure*}[h!]
\centering
\includegraphics[width=17.6cm,clip]{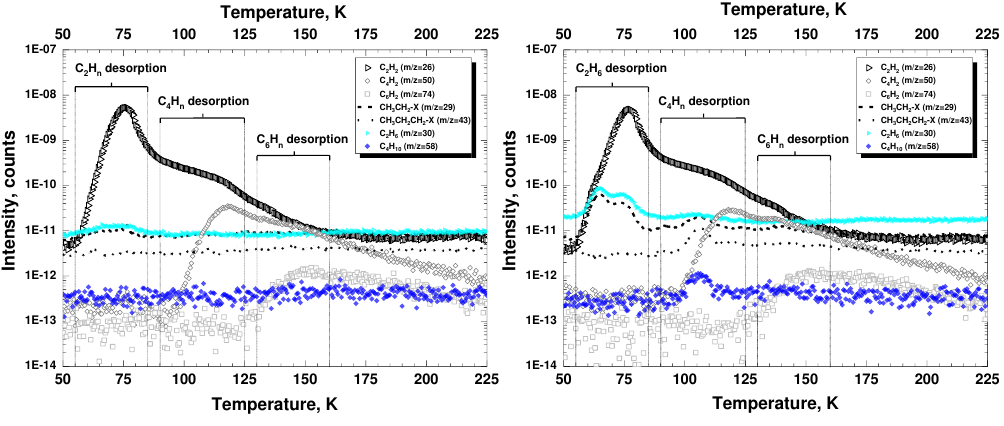}
\caption{Left:  QMS-TPD spectra in the range from 50 to 225 K for the selected m/z values obtained after the UV-exposure of 10 ML of pure HCCH ice with 4.6 $\times$ 10$^{16}$ photons cm$^{-2}$ at 10 K. Right:  QMS-TPD spectrum obtained after the hydrogenation of UV-exposed ice with 1.3 $\times$ 10$^{17}$ cm$^{-2}$ atoms at 10 K. The ordinate axis is presented as a logarithm of 10. An exponential decay of ion intensity appears as a straight line.}
\label{fig:1}
\end{figure*}

\subsection{Experimental methods}
The experiments are performed in the following procedure. First, a "pillow" of Ar ice (10 Langmuir\footnote{1 Langmuir (L) corresponds to exposure of substrate at 1 $\times$ 10$^{-6}$ torr pressure during one second at room temperature. This results in a surface coverage of about 1 monolayer. The usage of Langmuir estimation for the Ar ice is caused by the lack of absorption in the IR range.}) is deposited on top of a bare gold substrate. Then a $\sim$10 monolayer (ML) thick C$_2$H$_2$ ice layer is deposited on top of the predeposited Ar ice layer. The layered ice sample is exposed to the UV-photon beam generated by MDHL to achieve partial polymerization of C$_2$H$_2$ into its C$_{2n}$H$_2$ ($n>1$) polyynes derivatives. This polymerization has been extensively studied by \cite{Cuylle2014A&A} using a combination of mid-infrared and ultraviolet-visible (UV-VIS) spectroscopy \citep[see also][]{Compagnini2009, Wu2010ApJ, Puglisi2014NIMPB, Abplanalp2020ApJ}. The thickness of the grown C$_2$H$_2$ ice is well below the penetration depth of UV-photons. This aims to secure the uniform distribution of the formed products in the photo-processed ice. On the other hand, the interaction of UV-photons passing through the ice with the gold substrate may result in the production of photoelectrons. \cite{Hopkins1964BJAP} reported the work function of $>$4.7 eV for gold. The deposited ten Langmuir layer of Ar ice aims to prevent the interaction of the HCCH ice with the photoelectrons due to the limited penetration depth of photoelectrons in condensed phases \citep{Jo1991JChPh}.

The ice composition is monitored in situ by means of Fourier-transform reflection absorption infrared spectroscopy (RAIRS). Upon completion of the UV-exposure stage, a quadrupole mass spectrometer-temperature programmed desorption (QMS-TPD) measurement is performed. The substrate is rotated by 135$^{\circ}$ to face directly the ion source of the QMS. Then, during the TPD measurements, the ice is gently warmed up with a 2 K/min rate from 10 to 50 K to remove the "pillow" of Ar ice. Subsequently, the 5 K/min heating rate is used in the range of 50 to 250 K to increase the sensitivity of the applied TPD techniques. The sublimating species are continuously monitored by means of QMS. The formed species are then identified through the presence of their characteristic m/z (mass-to-charge) values at characteristic desorption temperatures. Thus, two independent parameters are used to identify the obtained products. 

The described experiment is then repeated using the very same experimental conditions followed by irradiation of the obtained ice with an overabundance of H atoms. Upon completion of the H-atom exposure stage, QMS-TPD measurement is performed by applying the same heating routine. The difference in the outcome of the performed experiments is used to identify the presence of four- and six-carbon bearing products of C$_{2n}$H$_2$ (n$>$1) polyynes hydrogenation, formed along with the expected products of well-investigated C$_2$H$_2$ hydrogenation \citep{Hiraoka2000ApJ, Kobayashi2017ApJ, Chuang2020A&A}.

Only the top monolayer of C$_2$H$_2$ ice is assumed to be available for hydrogenation by incident H-atoms \citep{Ioppolo2013RScI, Chuang2018ApJb}. This assumption is in agreement with the results of the pure HCCH ice hydrogenation experiment reported by \cite{Kobayashi2017ApJ}. The hydrogenation of 12 ML of pure HCCH ice at 10 K resulted in a relative HCCH consumption of 0.06 at an equilibrium state \cite[see Fig. 7 of][]{Kobayashi2017ApJ}.  This observation allows for the derivation of RAIRS setup-specific C$_2$H$_2$ band strength value. With this goal, pure C$_2$H$_2$ ice is hydrogenated upon reaching the maximum C$_2$H$_2$ consumption. Then, the difference in the total area for the selected C$_2$H$_2$ absorption features is obtained at the end of the experiment. This area corresponds to the absorption by $\sim$1 ML of C$_2$H$_2$ ice (where 1 ML surface coverage is assumed to be 1 $\times$ 10$^{15}$ cm$^{-2}$).  With the geometry utilized in our setup, the absorbance area equal to 0.028 cm$^{-1}$ was obtained for the C-H stretching mode of C$_2$H$_2$ ($\nu_3$, 3244 cm$^{-1}$). This corresponds to 0.064 cm$^{-1}$ in optical depth, resulting in the setup-specific RAIRS band strength values equal to 6.5 $\times$ 10$^{-17}$ cm molecule$^{-1}$.

\section{Results and discussion}\label{RESULTS AND DISCUSSIONS}
\subsection{Application of QMS-TPD technique}
The QMS-TPD spectra obtained after the UV-exposure of the deposited C$_2$H$_2$ ice are presented in the left panel of Fig. \ref{fig:1}. It is important to note that the ordinate axis is presented as a logarithm of 10 to highlight the appearance of the small desorption signals following the deviations from the exponential decays. Only the most relevant m/z values are selected and shown for better clarity. Three distinct TPD peaks corresponding to the desorption of three individual H-(C)$_{2n}$-H alkynes or polyynes can be clearly observed in the spectrum. These are C$_2$H$_2$ (m/z = 26), C$_4$H$_2$ (m/z = 50), and C$_6$H$_2$ (m/z = 74). Under our experimental conditions, the desorption of C$_2$H$_2$ starts at about 55 K and peaks at 75 K, while the desorption of C$_4$H$_2$ occurs at 95 K and peaks at 120 K. These temperature ranges are in agreement with the previous experimental results \citep[see][]{Zhou2009P&SS, Zhou2010ApJ, Abplanalp2017ApJ}. To our knowledge, no literature value has been reported for pure C$_6$H$_2$ ice. Under our experimental conditions, the corresponding m/z = 74 peak appears at 125 K and reaches the maximum at about 150 K. Several additional m/z values are presented in the spectra for reference. The m/z = 30 and m/z = 58 correspond to the "parent masses" of C$_2$H$_6$ and C$_4$H$_{10}$, i.e., the masses of ions obtained by the direct non-dissociative ionisation of C$_2$H$_6$ and C$_4$H$_{10}$, respectively. While the m/z = 29 and m/z=43 spectra correspond to the signals of C$_2$H$_5$$^+$ and C$_3$H$_7$$^+$, the two most likely ions obtained by dissociative ionization of various short linear aliphatic hydrocarbons in the ion head of the QMS. Appendix B in \cite{Schuhmann2019A&A}  offers typical examples of mass-spectra obtained for the short linear hydrocarbons. The QMS-TPD spectra presented in the left panel of Fig. \ref{fig:1} reveal the traces of C$_2$H$_6$ ice through the rise of m/z =30 and m/z = 29 signals in the range from 55 to 80 K. No signs of other saturated hydrocarbons can be found. 

\begin{figure}[h!]
\centering
\includegraphics[width=8.8cm,clip]{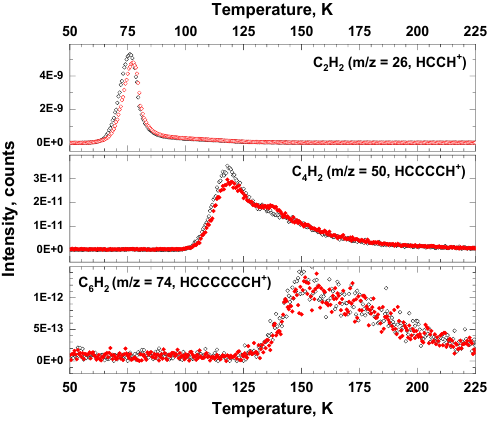}
\caption{Direct comparison between the fragments of QMS-TPD spectra in the range from 50 to 225 K obtained after the UV-exposure of 10 ML thick pure HCCH ice at 10 K with $>$4.6 $\times$ 10$^{16}$ cm$^{-2}$ photons (black empty circles) and the subsequent exposure of this photoprocessed ice with 1.3 $\times$ 10$^{17}$  cm$^{-2}$ H atoms at 10 K (red filled circles). The results obtained upon subtraction of the initial mass spectrum from the mass spectrum obtained upon hydrogenation of the ice for each of the presented m/z values are presented in Appendix \ref{appendixA}}
\label{fig:2}
\end{figure}

\begin{figure}[h!]
\centering
\includegraphics[width=8.8cm,clip]{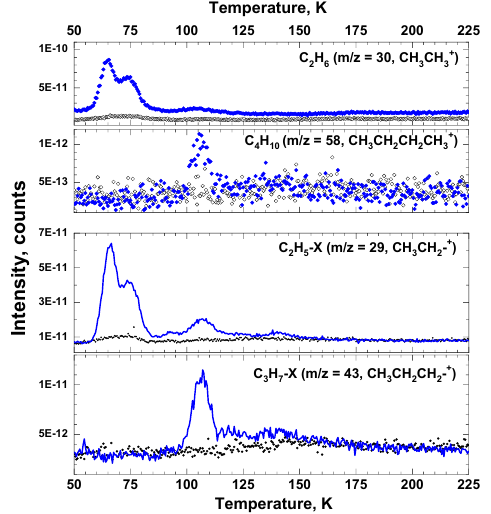}
\caption{Similar to Fig. \ref{fig:2}. M/z signals corresponding to linear saturated aliphatic hydrocarbons are shown.}
\label{fig:3}
\end{figure}

The right panel of Fig. \ref{fig:1} presents a general view of the QMS-TPD spectra obtained after the hydrogenation of the UV-exposed C$_2$H$_2$ ice under the same experimental conditions. Here, a minor drop in the intensity of C$_2$H$_2$, C$_4$H$_2$, and C$_6$H$_2$ desorption features can be distinguished (see Fig. \ref{fig:2} for more details). On the other hand, the appearance of new desorption features corresponding to the saturated aliphatic hydrocarbons can be observed. The rise of m/z = 30 (C$_2$H$_6$$^+$) and m/z = 29 (C$_2$H$_5$$^+$) signals in the temperature range between 55 and 85 K can be attributed to the desorption of C$_2$H$_6$. The rise of m/z = 58 (C$_4$H$_{10}$$^+$), m/z = 43 (C$_3$H$_7$$^+$) and m/z = 29 (C$_2$H$_5$$^+$) in the range from 95 to 115 K can be attributed to the desorption of C$_4$H$_{10}$. The third desorption feature in the range from 130 to 150 K can be tentatively assigned to semi-saturated or fully saturated hydrocarbons with six carbon atoms through the rise of m/z = 29 (C$_2$H$_5$$^+$).  More details are given in Fig. \ref{fig:3}.

The detailed comparison between the QMS-TPD curves obtained for the H-(C)$_{2n}$-H polyynes with and without the hydrogenation of the ice is presented in Fig. \ref{fig:2}. The linear scale is used for all of the plots for the direct comparison of the corresponding areas. The m/z signals assigned to C$_2$H$_2$ (HCCH, m/z = 26) and C$_4$H$_2$ (HCCCCH, m/z =50) demonstrate about 0.1 drop in the intensity upon hydrogenation of the ice. In the case of C$_6$H$_2$ (HCCCCCCH, m/z=74) molecules, the observation of such a drop can only be considered tentative because of the low signal-to-noise ratio. These drops in the areas of TPD curves in the right panel of Fig. \ref{fig:2} are due to the consumption of C$_2$H$_2$, C$_4$H$_2$ and C$_6$H$_2$ molecules in the upper layer of the UV-exposed ice through the reactions with H atoms. These experimental results show that the reactivity of unsaturated triple bonds of C$_4$H$_2$ (and, likely, C$_6$H$_2$) with H-atoms is similar to the reactivity of the C$_2$H$_2$ triple bond \cite[see also][]{Hiraoka2000ApJ, Kobayashi2017ApJ}. The shape and profile changes in the TPD curves obtained after the hydrogenation of the ice can be attributed to the more complex (non-uniform) composition of the ice. Indeed, the binding energies for polyynes are higher than those for their corresponding saturated counterparts \citep[see, e.g.,][]{Abplanalp2017ApJ}. This can be explained by the interactions between $\pi$-electrons of their triple bonds. This difference in the binding energies can affect the shape and profile of polyyne desorption features obtained for the mixed ice.

In addition to the data presented in Fig. \ref{fig:2}, we present the comparison between the QMS-TPD curves corresponding to the fully saturated aliphatic hydrocarbons obtained with and without H-atom exposure to the ice in Fig. \ref{fig:3}. The dissociative ionisation of various linear saturated aliphatic hydrocarbons in the ion source of the QMS results in the efficient cleavage of both C-C and C-H bonds. The upper panels in Fig. \ref{fig:3} show QMS-TPD curves obtained for the "parent masses" of C$_2$H$_6$ (CH$_3$CH$_3$$^+$, m/z = 30) and C$_4$H$_{10}$ (CH$_3$CH$_2$CH$_2$CH$_3$$^+$, m/z = 58). At the same time, the m/z = 29 (CH$_3$CH$_2$$^+$) and m/z = 43 (CH$_3$CH$_2$CH$_2$$^+$) correspond to the common fragments obtained after dissociative ionization of various linear aliphatic hydrocarbons. The peak in the range from 55 to 85 K can be confidently assigned to C$_2$H$_6$. The m/z = 29 to m/z = 30 signal ratio of 0.8 is consistent with the ratio reported for C$_2$H$_6$ in the NIST database\footnote{NIST Mass Spec Data Center, S. E. Stein, director, "Mass Spectra" in NIST Chemistry WebBook, NIST Standard Reference Database Number 69, eds. P. J. Linstrom and W. G. Mallard, National Institute of Standards and Technology, Gaithersburg, MD 20899, USA.}. The clear peaks for the characteristic m/z signals of linear C$_4$H$_{10}$ equal to 58, 43 and 29 are observed in the range from 90 to 105 K. These desorption temperatures are in close agreement with the values reported previously for C$_4$H$_{10}$ \citep[][the structural isomer is not specified]{Abplanalp2017ApJ}. In addition to the two resolved desorption peaks, a weaker desorption feature can be observed in the range from 125 to 150 K for m/z = 29 (CH$_3$CH$_2$$^+$) and m/z = 43 (CH$_3$CH$_2$CH$_2$$^+$). Observation of these dissociative ionization products is in agreement with the possible presence of C$_6$H$_{14}$ (CH$_3$CH$_2$CH$_2$CH$_2$CH$_2$CH$_3$); namely, a fully saturated derivative of C$_6$H$_2$ (HCCCCCCH). The observed desorption temperature range overlaps with that reported by \cite{Abplanalp2017ApJ} for various hydrocarbons consisting of six carbon atoms. However, without the identification of m/z = 86 “parent mass,” this attribution is considered tentative. The comparison between the observed intensities of m/z signals and the standard mass spectra available from the NIST database is presented in Appendix \ref{appendixB}.

\begin{figure}[h!]
\centering
\includegraphics[width=8.8cm,clip]{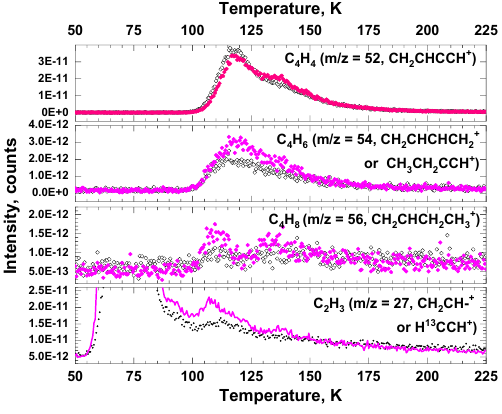}
\caption{Similar to Figs. \ref{fig:2} and \ref{fig:3}. M/z signals associated with the semi-saturated aliphatic hydrocarbons are presented.}
\label{fig:4}
\end{figure}

It should be noted that the m/z signals associated with the presence of semi-saturated hydrocarbons can also be identified in the mass spectra. However, the unambiguous identification of semi-saturated hydrocarbons becomes increasingly difficult due to the high number of possible structure isomers and the existing overlap between the m/z signals of fully saturated and semi-saturated hydrocarbons. The combination of C$_4$H$_4$, C$_4$H$_6$, and C$_4$H$_8$ alone has more than 20 stable structural isomers, excluding isotopes. Meanwhile, C$_6$H$_n$ semi-saturated hydrocarbons have more than 100 isomers. Some observations can be made based on the obtained laboratory data, see Fig. \ref{fig:4}. It is seen that prior to the hydrogenation of the UV-exposed ice, the peaks of m/z =52 (C$_4$H$_4$$^+$) and m/z = 54 (C$_4$H$_6$$^+$) signals are already present in the spectra. The desorption starts from 100 K and peaks in the range from 115 to 120 K. The m/z = 52 value corresponds to the “parent” and most intense signal of vinylacetylene (C$_4$H$_4$, CH$_2$CHCCH).  The desorption temperature registered for the m/z = 52 signal is higher than that observed for the m/z = 58 signal of C$_4$H$_{10}$ and is consistent with the literature value reported for the pure CH$_2$CHCCH ice by \cite{Kim2009ApJS} \citep[see Fig. \ref{fig:4} of][]{Kim2009ApJS}. Generally, a gradual shift towards lower desorption temperature is expected with the increase of hydrogenation degree, namely, T$_{\rm dep}$(C$_4$H$_2$) $>$ T$_{\rm dep}$(C$_4$H$_{4-8}$) $>$ T$_{\rm dep}$(C$_4$H$_{10}$), as, for example, described in \cite{Abplanalp2017ApJ}. The formation of CH$_2$CHCCH upon UV exposure of C$_2$H$_2$ ice is in agreement with the experimental results of \cite{Cuylle2014A&A}. The exact identification of produced semi-saturated hydrocarbons requires other experimental techniques and is beyond the scope of this work.

A weak desorption feature observed for the m/z = 54 (C$_4$H$_6$$^+$) signal can be attributed to the traces of one of the C$_4$H$_6$ isomers in the irradiated ice. The most probable candidates are ethylacetylene (CH$_3$CH$_2$CCH) and divinyl (CH$_2$CHCHCH$_2$). Here, the preference can be given to CH$_3$CH$_2$CCH. The observed m/z = 54 signal is the most intense peak in the mass spectra of CH$_3$CH$_2$CCH and corresponds to its non-dissociative ionization in the ion source of the QMS. Previous laboratory studies showed that surface hydrogenation of HCCH and HCCCH$_3$ triple bonds proceeds with a higher activation barrier than the hydrogenation of CH$_2$CH$_2$ and CH$_2$CHCH$_3$ double bonds, see \cite{Kobayashi2017ApJ} and \cite{Qasim2019ESCb}. Similar behavior can be expected for the hydrogenation of CH$_2$CHCCH triple and double bonds. Preferable formation of CH$_3$CH$_2$CCH can be suggested in this case. Although CH$_2$CHCHCH$_2$ is thermodynamically more stable than CH$_3$CH$_2$CCH due to the presence of the conjugated double bonds, the preference should be given to the kinetic argument. Only traces of the C$_4$H$_8$$^+$ m/z desorption feature can be observed in the QMS-TPD spectrum of UV-exposed ice prior to hydrogenation.

The m/z curves obtained after the hydrogenation of the ice reveal the overall consumption of C$_4$H$_4$ and the formation of C$_4$H$_6$ and C$_4$H$_8$. This observation is consistent with the increase of hydrocarbon hydrogenation degree following various H-atom addition reactions in the mixed ice. The simultaneous rise of m/z = 56 (C$_4$H$_8$$^+$) and m/z = 27 (CH$_2$CH$^+$) signals in the range from 100 to 115 K is consistent with the desorption of 1-butene (CH$_2$CHCH$_2$CH$_3$). The lower desorption temperature registered for m/z = 56 (C$_4$H$_8$$^+$) in comparison to m/z=52 (C$_4$H$_4$$^+$) and m/z=54 (C$_4$H$_6$$^+$) is in line with the higher hydrogenation degree of CH$_2$CHCH$_2$CH$_3$. It should be noted that the rise of m/z = 56 and m/z = 27 signals is also observed in the range from 125 to 150 K. In this range, the desorption of 6-carbon bearing hydrocarbons should occur, see \cite{Abplanalp2017ApJ}. Both of these m/z values are present in the mass spectra of fully saturated linear C$_6$H$_{14}$. 

\subsection{RAIRS and kinetic analysis}

\begin{figure*}[h!]
\centering
\includegraphics[width=17.6cm,clip]{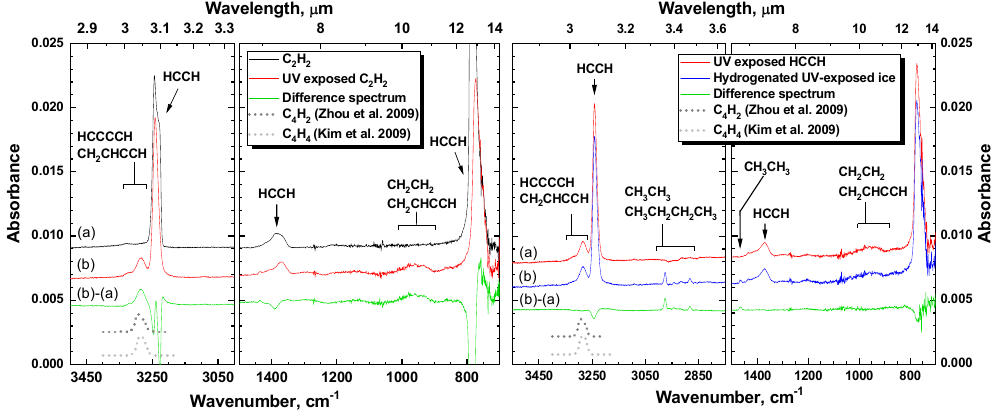}
\caption{Left: RAIR spectra obtained after the deposition of 10 ML of HCCH ice at 10 K (a, black line), after the exposure of this obtained ice with 4.6 $\times$ 10$^{16}$ cm$^{-2}$ photons (b, red line) and by subtracting the initial spectrum (a) from spectrum (b) (green line). Right: RAIR spectra obtained after the UV-exposure of 10 ML thick pure HCCH ice at 10 K with 4.6 $\times$ 10$^{16}$ cm$^{-2}$ photons (a, red line), the consequent exposure of this photoprocessed ice with 1.3 $\times$ 10$^{17}$ cm$^{-2}$ H atoms at 10 K (b, blue line), and by subtracting the initial spectrum (a) from spectrum (b) (green line). The grey dashed and dotted lines are the reference RAIRS spectra obtained for the pure C$_4$H$_2$ and C$_4$H$_4$ ices by \cite{Zhou2009P&SS} and \cite{Kim2009ApJS}.  All spectra are offset for clarity.}
\label{fig:5}
\end{figure*}

Complementarily to the QMS-TPD data, RAIRS results provide information about the ice composition at 10 K prior to the sample thermal processing and consequent destruction during the TPD experiment. Moreover, information about the phase changes and kinetic data can be acquired during the UV-processing and H-atom exposure of the ice. In the left panel of Fig. \ref{fig:5}, the RAIR spectra obtained after the deposition of pure C$_2$H$_2$ ice at 10 K on top of the 10 Langmuir Ar "pillow" (spectrum a), and the consequent UV-exposure of C$_2$H$_2$ ice (spectrum b) are presented. The obtained RAIR difference spectrum; namely, (b)-(a), is added for comparison. A clear decrease in the total area of the absorption features associated with the CH stretching mode ($\nu_3$, 3265-3210 cm$^{-1}$), CCH bending mode ($\nu_5$, 825-725 cm$^{-1}$), and the combinational mode ($\nu_4$+$\nu_5$, 1430-1350 cm$^{-1}$) is observed, indicating the consumption of HCCH. Moreover, the morphological changes in HCCH ice can be observed following the disappearance of C-H stretching and $\nu_4$+$\nu_5$ combinational mode absorption features associated with the contribution of "polycrystalline" HCCH ice (i.e., 3226 and 1390 cm$^{-1}$) and the corresponding growth of the C-H stretching mode absorption feature (i.e., 3240 cm$^{-1}$)  associated with "amorphous" HCCH ice \citep[see Fig. \ref{fig:8} of][]{Hudson2014Icar}. The "amorphous" and "polycrystalline" terms are used in quotes, as this band profile modification can also be attributed to the different orientation of HCCH molecules located in the interphase with vacuum or with the Ar ice "pillow", in comparison to the orientation of HCCH molecules in the bulk of HCCH ice. The effects in the interphase should impact only a few adjunct MLs of the ice. Thus, the contribution of the interphase is consistent with the low 10 ML thickness of the ices used in present study, in comparison to the thick $\rm {\mu}$m ice used by \cite{Hudson2014Icar}. A clear growth of the absorption feature in the range from 3308 to 3265 cm$^{-1}$ centered around 3285 cm$^{-1}$ is observed. This feature can be successfully assigned to the C-H stretching modes of both HCCCCH and CH$_2$CHCCH \citep{Torneng1980AcSpA, Zhou2009P&SS, Kim2009ApJS}, one of the most intensive modes for both of the species. The assignment of CH$_2$CHCCH can be further supported by the observation of a broad merged absorption feature in the range from 1020 to 900 cm$^{-1}$ with two peaks centered around 975 and 935 cm$^{-1}$. These features can be tentatively assigned to the $\nu_{14}$ and $\nu_{15}$ modes of CH$_2$CHCCH \citep{Torneng1980AcSpA, Kim2009ApJS}, along with the $\nu_7$ mode of C$_2$H$_4$. These observations further support the assignments done by the QMS-TPD technique, confirming the formation of both species already at 10 K.

\begin{figure}[h!]
\centering
\includegraphics[width=8.8cm,clip]{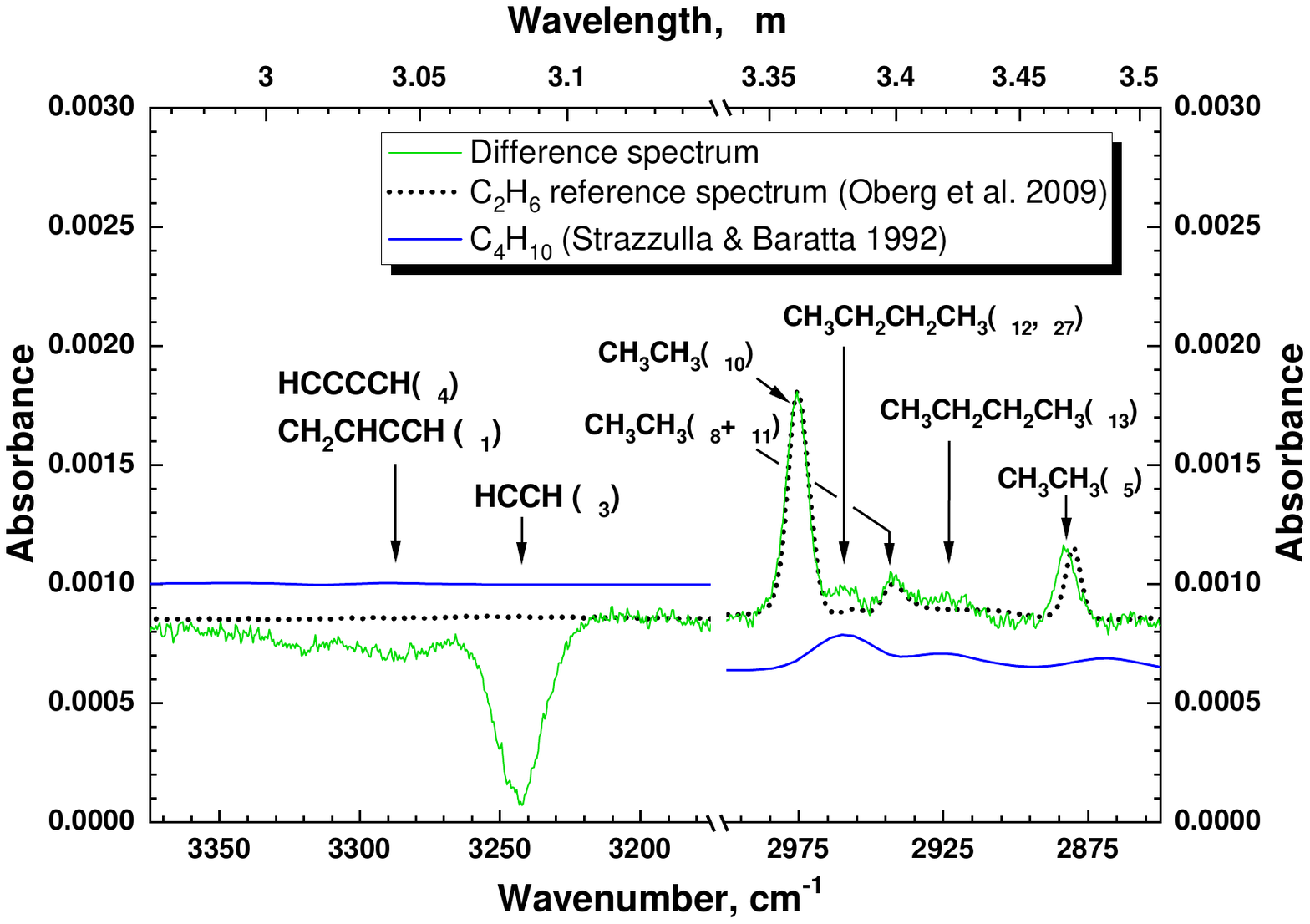}
\caption{Zoom-in into the CH stretching vibration modes region for the RAIR difference spectrum obtained after hydrogenation of UV-exposed C$_2$H$_2$ ice (green line). See the right panel of Fig. \ref{fig:5} for more details. The spectrum of pure C$_2$H$_6$ ice at 20 K is adapted from \cite{Öberg2009A&A} (black dots). The spectrum of pure C$_4$H$_{10}$ at 10 K is obtained by \cite{Straz1992A&A} (blue line). All spectra are offset for clarity.}
\label{fig:6}
\end{figure}

In the right panel of Fig. \ref{fig:5}, the RAIRS spectra (a) and (b) obtained prior to and after the H-atom exposure of the ice are presented. In a similar way, the RAIR difference spectrum (b)-(a) is shown for comparison. The expected decrease in the intensity of absorption features associated with unsaturated hydrocarbons (i.e., HCCH, HCCCCH and CH$_2$CHCCH) is observed. Simultaneously, multiple absorption features appear in the range from 2990 to 2870 cm$^{-1}$ and can be assigned to the C-H stretching modes of newly formed saturated hydrocarbons. The zoom-in spectral region of interest is presented in Fig. \ref{fig:6}. Three characteristic absorption features of CH$_3$CH$_3$ are observed at 2975, 2943, and 2882 cm$^{-1}$ corresponding to the $\nu_{10}$, $\nu_8$+$\nu_{11}$ and $\nu_5$ modes, respectively \citep{Comeford1961474, Kim2010ApJ, Turner2018ApJS}. However, in addition to the indicated absorption features of CH$_3$CH$_3$, the traces of two distinct absorption features of CH$_3$CH$_2$CH$_2$CH$_3$ can be tentatively identified at 2959 ($\nu_{12}$, $\nu_{27}$) and 2925 cm$^{-1}$ ($\nu_{13}$) corresponding to the asymmetric CH stretching modes of CH$_3$- and -CH$_2$- respectively \citep{Torneng1980AcSpA, Kim2010ApJ, Turner2018ApJS}. To further outline the plausible presence of -CH$_2$- asymmetric stretching mode of CH$_3$CH$_2$CH$_2$CH$_3$ the reference RAIRS spectrum of C$_2$H$_6$ obtained from \cite{Öberg2009A&A} is plotted versus the data obtained in the hydrogenation experiment. The asymmetric stretching mode of -CH$_2$- at 2925 cm$^{-1}$ can be considered a characteristic absorption feature for all saturated linear aliphatic hydrocarbons \citep[see, e.g., ][]{Yabuta2023Sci}. The RAIR difference spectrum obtained after hydrogenation of the ice does not reveal the growth of any absorption features that can be assigned to the semi-saturated hydrocarbons with C=C double bonds. 

\begin{figure}[h!]
\centering
\includegraphics[width=8.8cm,clip]{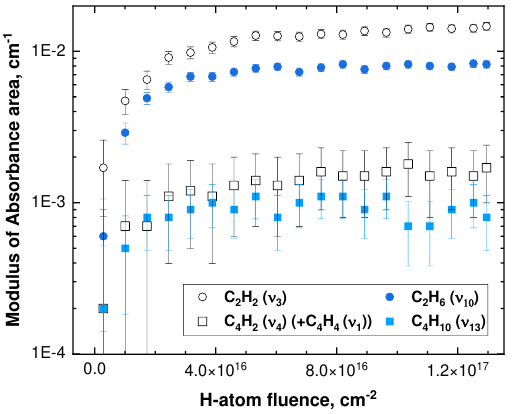}
\caption{Kinetic curves obtained for C$_2$H$_2$, C$_6$H$_2$, C$_4$H$_2$, and C$_4$H$_{10}$. Note: the area of the ($\nu_4$) absorption feature of C$_4$H$_2$ cannot be deconvoluted from the contribution of the ($\nu_1$) absorption feature of C$_4$H$_4$. Thus, the combined area is used for the integration. The modulus of absorbance is taken for C$_2$H$_2$ and C$_4$H$_2$ consumed in the experiments. This allows for the direct comparison with produced species. The ordinate axis is presented as a logarithm of 10.}
\label{fig:7}
\end{figure}

In first approximation the column density of the species is directly proportional to the total absorbance areas of their corresponding RAIRS absorption features. The "in situ" RAIRS technique allows for tracking the evolution of the total areas of the assigned absorbance features over the time of the H-exposure experiment. Figure \ref{fig:7} presents the obtained kinetic curves for the species identified in RAIR data. The total areas of absorption features assigned to the \(\ce{#CH}\)  stretching modes of \(\ce{HC#CH}\) (3265-3210 cm$^{-1}$) or \(\ce{HC#CC#CH}\) and \(\ce{CH$_2$=CHC#CH}\) (3308-3265 cm$^{-1}$) were used to track the consumption of these species following hydrogenation of their \(\ce{C#C}\) triple bonds. The formation of C$_2$H$_6$ is tracked through the absorption of its CH asymmetric stretching mode ($\nu_{10}$, 2995-2964 cm$^{-1}$). The integration of the peak (2929-2909 cm$^{-1}$) is used to track C$_4$H$_{10}$ following the tentative assignment of this absorption feature to the CH asymmetric stretch of -CH$_2$-. The modulus of absorbance is taken for C$_2$H$_2$ and C$_4$H$_2$ consumed in the experiments. This allows for the direct comparison with produced species. The presented kinetic curves show that the total areas of all selected features follow a very similar profile and reach the plateau around the same hydrogenation time. This indicates that the rates (or activation barriers) of H-atom addition reactions to the triple bonds of C$_2$H$_2$, C$_4$H$_2$, and C$_4$H$_4$ have similar values. Otherwise, a delay in the appearance of a plateau could be observed for the kinetic curve of a molecule with a slower H-atom addition rate.  The lack of shift between the C$_2$H$_2$ and C$_2$H$_6$ kinetic curves is observed. This is in agreement with the faster rate for hydrogenation of C=C double bonds of the intermediate C$_2$H$_4$ previously reported by \citep{Kobayashi2017ApJ}. The slower hydrogenation rate of the intermediate would result in a delay in the appearance of the fully saturated hydrocarbons with respect to the consumption of the initial reagent. Similarly, there is no shift between the C$_4$H$_2$ and C$_4$H$_{10}$ kinetic curves. Thus, a faster rate for hydrogenation of the intermediate C$_4$H$_{4-8}$ double bonds in comparison to the hydrogenation of C$_4$H$_2$ triple bonds can be inferred. This observation is also consistent with the lack of absorption features that can be assigned to the semi-saturated hydrocarbons with double bonds. The plateau appears in the range of H-atom fluence from 3 $\times$ 10$^{16}$ to 6 $\times$ 10$^{16}$ cm$^{-2}$. Assuming the H-atom flux in dark clouds is $\sim$10$^4$ atoms cm$^{-2}$ s$^{-1}$, these values of H-atom fluences are obtained at $\sim$1 $\times$ 10$^5$ and $\sim$2 $\times$ 10$^5$ years, respectively. It should be noted that these periods fall within the characteristic timescales of the dark clouds stage \citep{Boogert2015ARA&A,Chevance2020SSRv}.

\section{Astrochemical implications and conclusions}\label{ASTROCHEMICAL IMPLICATIONS AND CONCLUSIONS}
Our laboratory results show that fully saturated linear hydrocarbons (n-alkanes) such as n-C$_4$H$_{10}$ and plausibly n-C$_6$H$_{14}$ can be produced by the surface hydrogenation of the corresponding C$_{2n}$H$_2$ (H-(C)$_{2n}$-H) polyynes at 10 K. This work further extends the experimental finding obtained by \cite{Hiraoka2000ApJ}, \cite{Kobayashi2017ApJ} and \cite{Qasim2019ESCb}, which demonstrates the formation of C$_2$H$_4$, C$_2$H$_6$, C$_3$H$_6$, and C$_3$H$_8$ by surface hydrogenation of simplest alkynes C$_2$H$_2$ and C$_3$H$_8$, under the physical conditions resembling those founds in dark molecular clouds. \cite{Kobayashi2017ApJ} reported a four times higher rate for the hydrogenation of the HCCH triple bond than for the hydrogenation of CO molecules. This makes HCCH hydrogenation efficient at dark molecular cloud conditions. Comparable rates for the hydrogenation of HCCH, HCCCCH, and H$_2$CCHCCH triple bonds can be proposed based on the analysis of the presented kinetic data. Even higher rates are expected for hydrogenation of \(\ce{C=C}\) double bonds of intermediate semi-saturated hydrocarbons produced by hydrogenation of triple bonds \cite[see also][]{Kobayashi2017ApJ, Qasim2019ESCb}.

These experimental findings have a direct application to the surface chemistry at the early stages of star formation and, in particular, dark molecular clouds. This evolutionary stage is characterized by low pressures (10$^4$-10$^6$ cm$^{-3}$) and temperatures ($\sim$10 K). It sets up the initial chemical composition for the future evolutionary stages of star formation, providing material for the formation of protostars and protoplanetary disks. The data obtained by various space exploration missions such as Rosetta and Hayabusa2 suggests that part of the chemical inventory produced during the dark clouds stage may survive the harsh later stages of star formation and provide pristine material for comets and asteroids \citep{Rubin2015Sci, Yabuta2023Sci}. The chemistry of dark clouds is driven by the interplay between gas and dust. Ubiquitous sub-$\rm \mu$m dust grains provide the surface on which the simplest gas-phase chemical species (e.g., H, O, N, CO, etc.) accrete and react. Atom addition reactions among accreting species result in the formation of "H$_2$O-rich" and "CO-rich" ice mantles on the surface of the dust grains \citep{Oberg2011ApJ, Boogert2015ARA&A}.

Interstellar carbon chains and polyynes are efficiently produced early in dark molecular clouds through cold gas-phase reaction routes, see \cite{Taniguchi2024Ap&SS} and references therein.  The produced polyynes and carbon chains should accrete on the surface of dust grains and become available for reactions with other accreting species, such as H atoms. The efficient hydrogenation of polyynes demonstrated in the present study indicates that the formation of various linear aliphatic hydrocarbons can be expected at the beginning of the dark cloud stage. Moreover, the incorporation of other species (e.g., OH, O, N, etc.) into the polyyne hydrogenation chain will result in the formation of even more complex organic species \cite[see][]{Qasim2019A&Aa, Molpeceres2022A&A, Santos2024ESC, Chuang2020A&A, Chuang2021A&A, Chuang2024A&A}.

\begin{figure*}[h!]
\centering
\includegraphics[width=16.4cm,clip]{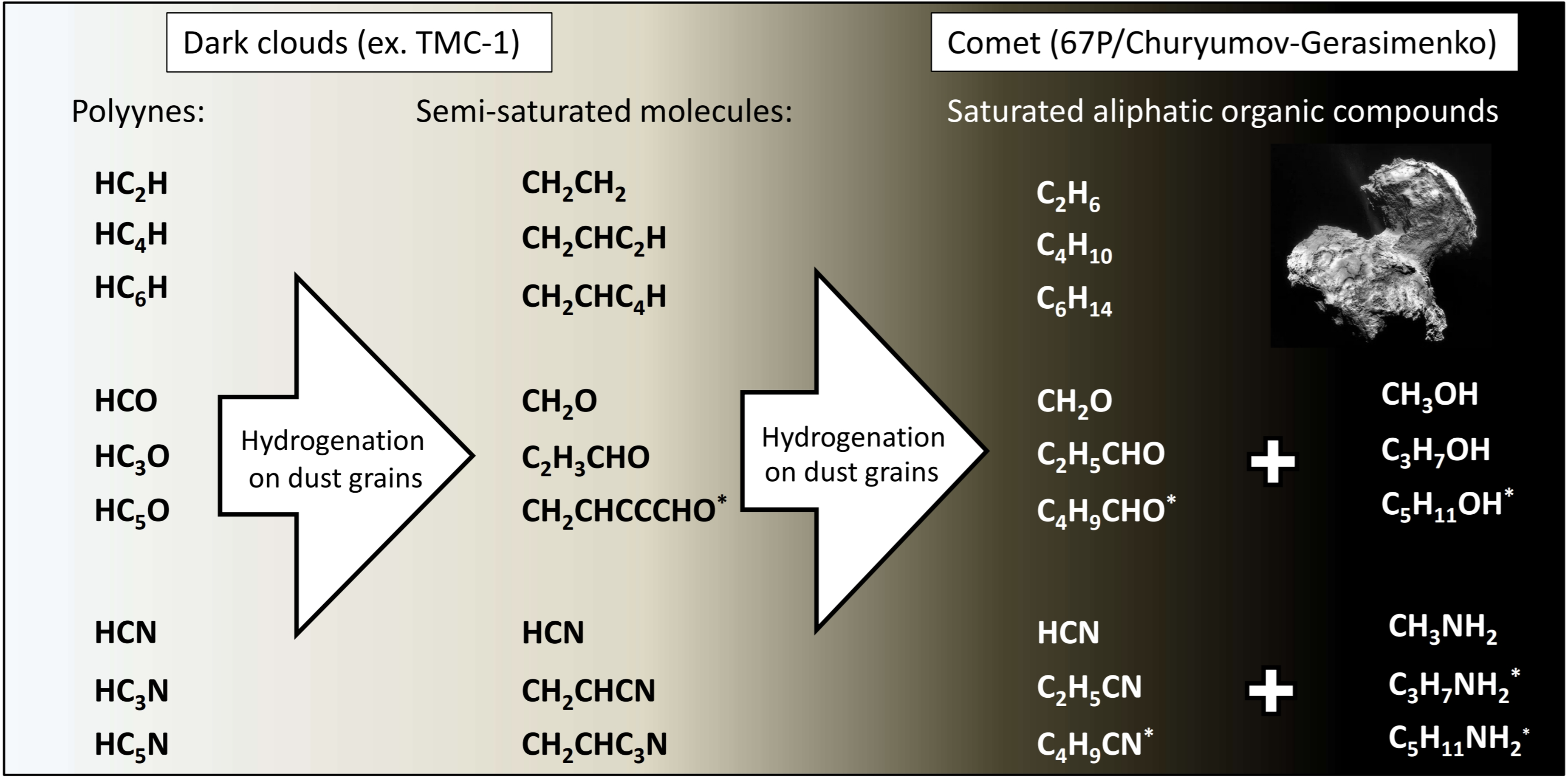}
\caption{Schematic representation of the overall suggested mechanism. Various polyynes observed in dark clouds (e.g., TMC-1) are presented in the left part of the reaction scheme.  The accretion of polyynes on the surface of the grains is followed by hydrogenation of their unsaturated CC triple bonds, resulting in the formation of their semi-saturated and fully saturated counterparts. Besides CH$_2$CHCCCHO, all of the first-generation products obtained by such hydrogenation have also been observed in the gas phase in TMC-1 or other dark clouds \protect\citep{Brown1981ApJ, Irvine1984OrLi, Agundez2021A&A, Cernicharo2001ApJ, Cernicharo2020A&A, Cernicharo2021A&Ab, Cernicharo2021A&Aa, Cernicharo2024A&A, Lee2021ApJ, McGuire2017ApJ}. In the right part of the reaction scheme, we present the fully saturated aliphatic organic compounds. Most of these molecules have been observed in the comet 67P/Churyumov-Gerasimenko. The heavy molecules whose identifications are still lacking are marked with asterisks. It should be noted that UV-photolysis and cosmic ray processing of fully saturated aliphatic compounds result in their efficient dehydrogenation. The high saturation degree of the detected aliphatic compounds hints at their partial formation through surface hydrogenation mechanisms \protect\citep{Altwegg2017MNRAS, Schuhmann2019A&A, Hanni2021A&A, Hanni2022NatCo, Hanni2023A&A}. The image of comet 67P/Churyumov-Gerasimenko is taken by ESA Rosetta on September 2, 2014.\protect\footnotemark}
\label{fig:8}
\end{figure*}

The formation of n-alkanes and semi-saturated aliphatic hydrocarbons early in dark clouds is important for the preservation of these species during the later stages of star formation. The "H$_2$O-rich" and "CO-rich" ice mantles along with coagulation of the icy grains can provide limited protection from UV-photons originating both from the cosmic ray induced UV-field and the protostar. Preservation of produced linear alkanes and the intermediate semi-saturated hydrocarbons followed by their integration into the pristine material of comets is consistent with the detection of alkanes released by comet 67P/Churyumov-Gerasimenko \cite[see][]{Schuhmann2019A&A, Altwegg2017MNRAS}. In addition,  \cite{Schuhmann2019A&A} reported that saturated alkanes account for 42\% of the total amount of detected hydrocarbons in Comet 67P/Churyumov-Gerasimenko, while the contribution of unsaturated hydrocarbons is 56\%. The cycloalkanes contribution is calculated to be at most 2 \%. A minimum contribution of cycloalkanes is consistent with the production of aliphatic hydrocarbons by the hydrogenation of linear polyynes. It should be noted that even though the reference spectra of linear alkanes were used in the study by \cite{Schuhmann2019A&A}, the contribution of branched alkanes could not be deconvoluted in that work. 

Besides observations in comet 67P, hydrocarbons have been recently identified on the surface of Ceres \citep{DeSanctis2017Sci} and in the samples from the asteroid Ruygu delivered by the Hayabusa2 mission. Observation of asymmetric stretching modes of CH$_3$- (2960 cm$^{-1}$, 3.38 $\rm {\mu}$m) and -CH$_2$- (2930 cm$^{-1}$, 3.41 $\rm {\mu}$m) along with the other bands in the infrared range is used to constrain the presence of aliphatic hydrocarbons in Ryugu samples by \cite{Yabuta2023Sci}. Identification of characteristic -CH$_2$- modes is consistent with the presence of linear alkanes comprising three or more carbon atoms or with the presence of linearly saturated hydrocarbon segments incorporated into the organic residue. Moreover, \cite{Yabuta2023Sci} report a relatively high -CH$_2$- to -CH$_3$ peak intensity ratio equal to 1.9. Based on these observations and on the comparison with the ratios observed in Murchison and Ivuna meteorites, \cite{Yabuta2023Sci} inferred that Ryugu’s organic matter may contain longer aliphatic chains than in the meteoritic insoluble organic matters (IOMs). 

Gas-phase observations of saturated linear hydrocarbons are challenging because of the low dipole moment of these molecules. Alternative method for alkanes identification in the ISM can be suggested. If alkanes are present on the grain surface early during the dark cloud stage, these species can become a target for JWST observations in the mid-infrared range. Albeit the abundance of each individual alkane is expected to be low, their absorption features overlap \citep{Torneng1980AcSpA, Turner2018ApJS}. This will result in the cumulative add-up of individual absorption of each species into the combined optical depth. In this way the whole class of chemical compounds can be observed simultaneously. Besides the CH$_3$- (2960 cm$^{-1}$, 3.38 $\rm {\mu}$m) and -CH$_2$- (2930 cm$^{-1}$, 3.41 $\rm {\mu}$m) bands mentioned earlier, the CH$_3$-/-CH$_2$- deformation/scissoring modes at 1460 cm$^{-1}$ (6.85 $\rm {\mu}$m) and CH$_3$- symmetric deformation mode at 1380 cm$^{-1}$ (7.25 $\rm {\mu}$m) are good potential candidates for such "family" spectral indicators. 

Semi-saturated hydrocarbons can be observed in the gas phase. Such observations would require a mechanism responsible for the partial grain-to-gas transfer of produced intermediate species at low temperatures, such as reactive desorption \cite[see, e.g.,][]{Minissale2016MNRAS, Oba2018NatAs, Chuang2018ApJb}. \cite{Cernicharo2021A&Aa} reported the first detection of C$_4$H$_4$ (vinylacetylene) in TMC-1 through observation of its 6 transitions. C$_4$H$_6$ (ethylacetylene) was also tentatively identified in their study. So far, there have been no reports about successful detections of 1-butene, despite the fact that its structural isomer isobutene was observed in TMC-1; see \cite{Fatima2023A&A}. If solid-state formation routes contribute to the gas-phase enrichment of these species, then other semi-saturated hydrocarbons can be set as targets for future gas-phase observations. The precise identification of intermediate four- and six-carbon bearing semi-saturated hydrocarbons in the present work was challenging. This is mainly because of the high number of possible isomers and an overlap between their characteristic m/z values in the mass spectra. The determination of the exact C$_4$H$_2$ and C$_6$H$_2$ hydrogenation sequences could serve as the subject of future experimental and theoretical studies. Nevertheless, we suggest that the preference should be given to the observations of semi-saturated hydrocarbons with triple bonds due to their higher stability.

\footnotetext{Emily Lakdawalla, Comet jets! NavCam view of comet Churyumov-Gerasimenko on September 2, 2014, ESA, Rosetta, https://www.planetary.org/space-images/comet-cg-20140902-jets}

Finally, the present work uses the two-step experimental technique to investigate the polyynes reactions with H atoms. The mixture of unstable polyynes was synthesized in situ on the substrate in the main UHV chamber of the setup by UV-irradiation of C$_2$H$_2$ ice. Then, the hydrogenation process of the obtained mixed ice using an H-atom beam was performed. Simultaneous hydrogenation of the mixed ice allowed us to compare the relative reactivity of involved species. This approach can be used to study the hydrogenation of other species, which might not otherwise be so easily obtained in the lab. This may include cyanopolyynes or oxygen- and sulphur-bearing carbon chains. The former, for example, can be synthesized by UV-exposure to HCN/C$_2$H$_2$ mixed ice. Then, their overall reactivity with H-atoms can be investigated. Analogously to  the present study and the works of \cite{Kobayashi2017ApJ} and \cite{Qasim2019ESCb}, we can expect that \(\ce{C#C}\) triple bonds of cyanopolyynes can participate in hydrogenation reactions that result in the formation of aliphatic nitriles or even amines \cite[see also][]{Theule2011A&A, Nguyen2019A&A}.  Similarly, the hydrogenation of (H)C$_n$O and C$_n$S carbon chains can result in the formation of various aliphatic aldehydes, alcohols, and thiols. We give more details in Fig. \ref{fig:8}. These substituted hydrocarbons have direct astrobiological implications as they are amphiphilic and may play a part in the formation of micelles in aqueous solutions, the most primitive prototypes of cells and organelles \citep{Segre2001OLEB, TREVORS2001573, Stevenson2015SciA}.

\begin{acknowledgements}
     This work is dedicated to the memory of Prof. dr. Harold V. J. Linnartz. This research was funded through financial support from the European Union’s Horizon 2020 research and innovation programme under the Marie Skłodowska-Curie actions grant agreement no. 664931, and the Netherlands Research School for Astronomy (NOVA). GF thanks the Xinjiang Tianchi Talent Program (2024). GAB and MEP acknowledge Progetto Premiale 2012 ‘iALMA’ grant (CUP C52I13000140001) and INAF Grant “Ricerca Fondamentale 2022” (CUP C63C22000900005). KJC is grateful for support from NWO via a VENI fellowship (VI.Veni.212.296). We would like to thanks Juan Tuo for her help in preparation of this manuscript. 
\end{acknowledgements}

%-------------------------------------------------------------------
  \bibliographystyle{aa} % style aa.bst
  \bibliography{aa} % your references Yourfile.bib

\begin{thebibliography}{94}
\expandafter\ifx\csname natexlab\endcsname\relax\def\natexlab#1{#1}\fi

\bibitem[{{Abplanalp} \& {Kaiser}(2017)}]{Abplanalp2017ApJ}
{Abplanalp}, M.~J. \& {Kaiser}, R.~I. 2017, \apj, 836, 195

\bibitem[{{Abplanalp} \& {Kaiser}(2020)}]{Abplanalp2020ApJ}
{Abplanalp}, M.~J. \& {Kaiser}, R.~I. 2020, \apj, 889, 3

\bibitem[{{Ag{\'u}ndez} {et~al.}(2021){Ag{\'u}ndez}, {Marcelino}, {Tercero}, {Cabezas}, {de Vicente}, \& {Cernicharo}}]{Agundez2021A&A}
{Ag{\'u}ndez}, M., {Marcelino}, N., {Tercero}, B., {et~al.} 2021, \aap, 649, L4

\bibitem[{{Altwegg} {et~al.}(2017){Altwegg}, {Balsiger}, {Berthelier}, {Bieler}, {Calmonte}, {Fuselier}, {Goesmann}, {Gasc}, {Gombosi}, {Le Roy}, {de Keyser}, {Morse}, {Rubin}, {Schuhmann}, {Taylor}, {Tzou}, \& {Wright}}]{Altwegg2017MNRAS}
{Altwegg}, K., {Balsiger}, H., {Berthelier}, J.~J., {et~al.} 2017, \mnras, 469, S130

\bibitem[{{Belloche} {et~al.}(2022){Belloche}, {Garrod}, {Zingsheim}, {M{\"u}ller}, \& {Menten}}]{Belloche2022A&A}
{Belloche}, A., {Garrod}, R.~T., {Zingsheim}, O., {M{\"u}ller}, H.~S.~P., \& {Menten}, K.~M. 2022, \aap, 662, A110

\bibitem[{{Bergner} {et~al.}(2018){Bergner}, {Guzm{\'a}n}, {{\"O}berg}, {Loomis}, \& {Pegues}}]{Bergner2018ApJ}
{Bergner}, J.~B., {Guzm{\'a}n}, V.~G., {{\"O}berg}, K.~I., {Loomis}, R.~A., \& {Pegues}, J. 2018, \apj, 857, 69

\bibitem[{{Boogert} {et~al.}(2015){Boogert}, {Gerakines}, \& {Whittet}}]{Boogert2015ARA&A}
{Boogert}, A.~C.~A., {Gerakines}, P.~A., \& {Whittet}, D. C.~B. 2015, \araa, 53, 541

\bibitem[{{Brown}(1981)}]{Brown1981ApJ}
{Brown}, R.~L. 1981, \apjl, 248, L119

\bibitem[{{Butscher} {et~al.}(2016){Butscher}, {Duvernay}, {Danger}, \& {Chiavassa}}]{Butscher2016A&A}
{Butscher}, T., {Duvernay}, F., {Danger}, G., \& {Chiavassa}, T. 2016, \aap, 593, A60

\bibitem[{{Butscher} {et~al.}(2017){Butscher}, {Duvernay}, {Rimola}, {Segado-Centellas}, \& {Chiavassa}}]{Butscher2017PCCP}
{Butscher}, T., {Duvernay}, F., {Rimola}, A., {Segado-Centellas}, M., \& {Chiavassa}, T. 2017, Physical Chemistry Chemical Physics (Incorporating Faraday Transactions), 19, 2857

\bibitem[{{Butscher} {et~al.}(2015){Butscher}, {Duvernay}, {Theule}, {Danger}, {Carissan}, {Hagebaum-Reignier}, \& {Chiavassa}}]{Butscher2015MNRAS}
{Butscher}, T., {Duvernay}, F., {Theule}, P., {et~al.} 2015, \mnras, 453, 1587

\bibitem[{{Cernicharo} {et~al.}(2021{\natexlab{a}}){Cernicharo}, {Ag{\'u}ndez}, {Cabezas}, {Tercero}, {Marcelino}, {Fuentetaja}, {Pardo}, \& {de Vicente}}]{Cernicharo2021A&Ab}
{Cernicharo}, J., {Ag{\'u}ndez}, M., {Cabezas}, C., {et~al.} 2021{\natexlab{a}}, \aap, 656, L21

\bibitem[{{Cernicharo} {et~al.}(2021{\natexlab{b}}){Cernicharo}, {Cabezas}, {Ag{\'u}ndez}, {Tercero}, {Pardo}, {Marcelino}, {Gallego}, {Tercero}, {L{\'o}pez-P{\'e}rez}, \& {de Vicente}}]{Cernicharo2021A&Aa}
{Cernicharo}, J., {Cabezas}, C., {Ag{\'u}ndez}, M., {et~al.} 2021{\natexlab{b}}, \aap, 648, L3

\bibitem[{{Cernicharo} {et~al.}(2022){Cernicharo}, {Fuentetaja}, {Cabezas}, {Ag{\'u}ndez}, {Marcelino}, {Tercero}, {Pardo}, \& {de Vicente}}]{Cernicharo2022A&A}
{Cernicharo}, J., {Fuentetaja}, R., {Cabezas}, C., {et~al.} 2022, \aap, 663, L5

\bibitem[{{Cernicharo} {et~al.}(2001){Cernicharo}, {Heras}, {Tielens}, {Pardo}, {Herpin}, {Gu{\'e}lin}, \& {Waters}}]{Cernicharo2001ApJ}
{Cernicharo}, J., {Heras}, A.~M., {Tielens}, A.~G.~G.~M., {et~al.} 2001, \apjl, 546, L123

\bibitem[{{Cernicharo} {et~al.}(2020){Cernicharo}, {Marcelino}, {Ag{\'u}ndez}, {Berm{\'u}dez}, {Cabezas}, {Tercero}, \& {Pardo}}]{Cernicharo2020A&A}
{Cernicharo}, J., {Marcelino}, N., {Ag{\'u}ndez}, M., {et~al.} 2020, \aap, 642, L8

\bibitem[{{Cernicharo} {et~al.}(2024){Cernicharo}, {Tercero}, {Ag{\'u}ndez}, {Cabezas}, {Fuentetaja}, {Marcelino}, \& {de Vicente}}]{Cernicharo2024A&A}
{Cernicharo}, J., {Tercero}, B., {Ag{\'u}ndez}, M., {et~al.} 2024, \aap, 686, A139

\bibitem[{{Chen} {et~al.}(2014){Chen}, {Chuang}, {Mu{\~n}oz Caro}, {Nuevo}, {Chu}, {Yih}, {Ip}, \& {Wu}}]{Chen2014ApJ}
{Chen}, Y.~J., {Chuang}, K.~J., {Mu{\~n}oz Caro}, G.~M., {et~al.} 2014, \apj, 781, 15

\bibitem[{{Chevance} {et~al.}(2020){Chevance}, {Kruijssen}, {Vazquez-Semadeni}, {Nakamura}, {Klessen}, {Ballesteros-Paredes}, {Inutsuka}, {Adamo}, \& {Hennebelle}}]{Chevance2020SSRv}
{Chevance}, M., {Kruijssen}, J.~M.~D., {Vazquez-Semadeni}, E., {et~al.} 2020, \ssr, 216, 50

\bibitem[{{Chuang} {et~al.}(2020){Chuang}, {Fedoseev}, {Qasim}, {Ioppolo}, {J{\"a}ger}, {Henning}, {Palumbo}, {van Dishoeck}, \& {Linnartz}}]{Chuang2020A&A}
{Chuang}, K.~J., {Fedoseev}, G., {Qasim}, D., {et~al.} 2020, \aap, 635, A199

\bibitem[{{Chuang} {et~al.}(2017){Chuang}, {Fedoseev}, {Qasim}, {Ioppolo}, {van Dishoeck}, \& {Linnartz}}]{Chuang2017MNRAS}
{Chuang}, K.~J., {Fedoseev}, G., {Qasim}, D., {et~al.} 2017, \mnras, 467, 2552

\bibitem[{{Chuang} {et~al.}(2018{\natexlab{a}}){Chuang}, {Fedoseev}, {Qasim}, {Ioppolo}, {van Dishoeck}, \& {Linnartz}}]{Chuang2018A&Aa}
{Chuang}, K.~J., {Fedoseev}, G., {Qasim}, D., {et~al.} 2018{\natexlab{a}}, \aap, 617, A87

\bibitem[{{Chuang} {et~al.}(2018{\natexlab{b}}){Chuang}, {Fedoseev}, {Qasim}, {Ioppolo}, {van Dishoeck}, \& {Linnartz}}]{Chuang2018ApJb}
{Chuang}, K.~J., {Fedoseev}, G., {Qasim}, D., {et~al.} 2018{\natexlab{b}}, \apj, 853, 102

\bibitem[{{Chuang} {et~al.}(2021){Chuang}, {Fedoseev}, {Scir{\`e}}, {Baratta}, {J{\"a}ger}, {Henning}, {Linnartz}, \& {Palumbo}}]{Chuang2021A&A}
{Chuang}, K.~J., {Fedoseev}, G., {Scir{\`e}}, C., {et~al.} 2021, \aap, 650, A85

\bibitem[{{Chuang} {et~al.}(2024){Chuang}, {J{\"a}ger}, {Santos}, \& {Henning}}]{Chuang2024A&A}
{Chuang}, K.~J., {J{\"a}ger}, C., {Santos}, J.~C., \& {Henning}, T. 2024, \aap, 687, A7

\bibitem[{Comeford \& Gould(1961)}]{Comeford1961474}
Comeford, J. \& Gould, J.~H. 1961, Journal of Molecular Spectroscopy, 5, 474

\bibitem[{Compagnini {et~al.}(2009)Compagnini, D’Urso, Puglisi, Baratta, \& Strazzulla}]{Compagnini2009}
Compagnini, G., D’Urso, L., Puglisi, O., Baratta, G., \& Strazzulla, G. 2009, Carbon, 47, 1605

\bibitem[{{Cuadrado} {et~al.}(2015){Cuadrado}, {Goicoechea}, {Pilleri}, {Cernicharo}, {Fuente}, \& {Joblin}}]{Cuadrado2015A&A}
{Cuadrado}, S., {Goicoechea}, J.~R., {Pilleri}, P., {et~al.} 2015, \aap, 575, A82

\bibitem[{{Cuylle} {et~al.}(2014){Cuylle}, {Zhao}, {Strazzulla}, \& {Linnartz}}]{Cuylle2014A&A}
{Cuylle}, S.~H., {Zhao}, D., {Strazzulla}, G., \& {Linnartz}, H. 2014, \aap, 570, A83

\bibitem[{{De Sanctis} {et~al.}(2017){De Sanctis}, {Ammannito}, {McSween}, {Raponi}, {Marchi}, {Capaccioni}, {Capria}, {Carrozzo}, {Ciarniello}, {Fonte}, {Formisano}, {Frigeri}, {Giardino}, {Longobardo}, {Magni}, {McFadden}, {Palomba}, {Pieters}, {Tosi}, {Zambon}, {Raymond}, \& {Russell}}]{DeSanctis2017Sci}
{De Sanctis}, M.~C., {Ammannito}, E., {McSween}, H.~Y., {et~al.} 2017, Science, 355, 719

\bibitem[{{Fatima} {et~al.}(2023){Fatima}, {M{\"u}ller}, {Zingsheim}, {Lewen}, {Rivilla}, {Jim{\'e}nez-Serra}, {Mart{\'\i}n-Pintado}, \& {Schlemmer}}]{Fatima2023A&A}
{Fatima}, M., {M{\"u}ller}, H. S.~P., {Zingsheim}, O., {et~al.} 2023, \aap, 680, A25

\bibitem[{{Fedoseev} {et~al.}(2017){Fedoseev}, {Chuang}, {Ioppolo}, {Qasim}, {van Dishoeck}, \& {Linnartz}}]{Fedoseev2017ApJ}
{Fedoseev}, G., {Chuang}, K.~J., {Ioppolo}, S., {et~al.} 2017, \apj, 842, 52

\bibitem[{{Fedoseev} {et~al.}(2016){Fedoseev}, {Chuang}, {van Dishoeck}, {Ioppolo}, \& {Linnartz}}]{Fedoseev2016MNRAS}
{Fedoseev}, G., {Chuang}, K.~J., {van Dishoeck}, E.~F., {Ioppolo}, S., \& {Linnartz}, H. 2016, \mnras, 460, 4297

\bibitem[{{Fedoseev} {et~al.}(2015){Fedoseev}, {Cuppen}, {Ioppolo}, {Lamberts}, \& {Linnartz}}]{Fedoseev2015MNRAS}
{Fedoseev}, G., {Cuppen}, H.~M., {Ioppolo}, S., {Lamberts}, T., \& {Linnartz}, H. 2015, \mnras, 448, 1288

\bibitem[{{Fedoseev} {et~al.}(2022){Fedoseev}, {Qasim}, {Chuang}, {Ioppolo}, {Lamberts}, {van Dishoeck}, \& {Linnartz}}]{Fedoseev2022ApJ}
{Fedoseev}, G., {Qasim}, D., {Chuang}, K.-J., {et~al.} 2022, \apj, 924, 110

\bibitem[{{Fuentetaja} {et~al.}(2022{\natexlab{a}}){Fuentetaja}, {Ag{\'u}ndez}, {Cabezas}, {Tercero}, {Marcelino}, {Pardo}, {de Vicente}, \& {Cernicharo}}]{Fuentetaja2022A&Ab}
{Fuentetaja}, R., {Ag{\'u}ndez}, M., {Cabezas}, C., {et~al.} 2022{\natexlab{a}}, \aap, 667, L4

\bibitem[{{Fuentetaja} {et~al.}(2022{\natexlab{b}}){Fuentetaja}, {Cabezas}, {Ag{\'u}ndez}, {Tercero}, {Marcelino}, {Pardo}, {de Vicente}, \& {Cernicharo}}]{Fuentetaja2022A&Aa}
{Fuentetaja}, R., {Cabezas}, C., {Ag{\'u}ndez}, M., {et~al.} 2022{\natexlab{b}}, \aap, 663, L3

\bibitem[{{Gerakines} \& {Moore}(2001)}]{Gerakines2001Icar}
{Gerakines}, P.~A. \& {Moore}, M.~H. 2001, \icarus, 154, 372

\bibitem[{{Guzm{\'a}n} {et~al.}(2021){Guzm{\'a}n}, {Bergner}, {Law}, {{\"O}berg}, {Walsh}, {Cataldi}, {Aikawa}, {Bergin}, {Czekala}, {Huang}, {Andrews}, {Loomis}, {Zhang}, {Le Gal}, {Alarc{\'o}n}, {Ilee}, {Teague}, {Cleeves}, {Wilner}, {Long}, {Schwarz}, {Bosman}, {P{\'e}rez}, {M{\'e}nard}, \& {Liu}}]{Guzman2021ApJS}
{Guzm{\'a}n}, V.~V., {Bergner}, J.~B., {Law}, C.~J., {et~al.} 2021, \apjs, 257, 6

\bibitem[{{H{\"a}nni} {et~al.}(2023){H{\"a}nni}, {Altwegg}, {Baklouti}, {Combi}, {Fuselier}, {De Keyser}, {M{\"u}ller}, {Rubin}, \& {Wampfler}}]{Hanni2023A&A}
{H{\"a}nni}, N., {Altwegg}, K., {Baklouti}, D., {et~al.} 2023, \aap, 678, A22

\bibitem[{{H{\"a}nni} {et~al.}(2021){H{\"a}nni}, {Altwegg}, {Balsiger}, {Combi}, {Fuselier}, {De Keyser}, {Pestoni}, {Rubin}, \& {Wampfler}}]{Hanni2021A&A}
{H{\"a}nni}, N., {Altwegg}, K., {Balsiger}, H., {et~al.} 2021, \aap, 647, A22

\bibitem[{{H{\"a}nni} {et~al.}(2022){H{\"a}nni}, {Altwegg}, {Combi}, {Fuselier}, {De Keyser}, {Rubin}, \& {Wampfler}}]{Hanni2022NatCo}
{H{\"a}nni}, N., {Altwegg}, K., {Combi}, M., {et~al.} 2022, Nature Communications, 13, 3639

\bibitem[{{Hiraoka} {et~al.}(2000){Hiraoka}, {Takayama}, {Euchi}, {Handa}, \& {Sato}}]{Hiraoka2000ApJ}
{Hiraoka}, K., {Takayama}, T., {Euchi}, A., {Handa}, H., \& {Sato}, T. 2000, \apj, 532, 1029

\bibitem[{{Hopkins} \& {Riviere}(1964)}]{Hopkins1964BJAP}
{Hopkins}, B.~J. \& {Riviere}, J.~C. 1964, British Journal of Applied Physics, 15, 941

\bibitem[{{Hudson} {et~al.}(2014){Hudson}, {Ferrante}, \& {Moore}}]{Hudson2014Icar}
{Hudson}, R.~L., {Ferrante}, R.~F., \& {Moore}, M.~H. 2014, \icarus, 228, 276

\bibitem[{{Ilee} {et~al.}(2021){Ilee}, {Walsh}, {Booth}, {Aikawa}, {Andrews}, {Bae}, {Bergin}, {Bergner}, {Bosman}, {Cataldi}, {Cleeves}, {Czekala}, {Guzm{\'a}n}, {Huang}, {Law}, {Le Gal}, {Loomis}, {M{\'e}nard}, {Nomura}, {{\"O}berg}, {Qi}, {Schwarz}, {Teague}, {Tsukagoshi}, {Wilner}, {Yamato}, \& {Zhang}}]{Ilee2021ApJS}
{Ilee}, J.~D., {Walsh}, C., {Booth}, A.~S., {et~al.} 2021, \apjs, 257, 9

\bibitem[{{Ioppolo} {et~al.}(2013){Ioppolo}, {Fedoseev}, {Lamberts}, {Romanzin}, \& {Linnartz}}]{Ioppolo2013RScI}
{Ioppolo}, S., {Fedoseev}, G., {Lamberts}, T., {Romanzin}, C., \& {Linnartz}, H. 2013, Review of Scientific Instruments, 84, 073112

\bibitem[{{Irvine} \& {Hjalmarson}(1984)}]{Irvine1984OrLi}
{Irvine}, W.~M. \& {Hjalmarson}, A. 1984, Origins of Life, 14, 15

\bibitem[{{Jim{\'e}nez-Serra} {et~al.}(2022){Jim{\'e}nez-Serra}, {Rodr{\'\i}guez-Almeida}, {Mart{\'\i}n-Pintado}, {Rivilla}, {Melosso}, {Zeng}, {Colzi}, {Kawashima}, {Hirota}, {Puzzarini}, {Tercero}, {de Vicente}, {Rico-Villas}, {Requena-Torres}, \& {Mart{\'\i}n}}]{Jimenez-Serra2022A&A}
{Jim{\'e}nez-Serra}, I., {Rodr{\'\i}guez-Almeida}, L.~F., {Mart{\'\i}n-Pintado}, J., {et~al.} 2022, \aap, 663, A181

\bibitem[{{Jo} \& {White}(1991)}]{Jo1991JChPh}
{Jo}, S.~K. \& {White}, J.~M. 1991, \jcp, 94, 5761

\bibitem[{{Khlifi} {et~al.}(1995){Khlifi}, {Paillous}, {Delpech}, {Nishio}, {Bruston}, \& {Raulin}}]{Khlifi1995JMoSp}
{Khlifi}, M., {Paillous}, P., {Delpech}, C., {et~al.} 1995, Journal of Molecular Spectroscopy, 174, 116

\bibitem[{{Kim} {et~al.}(2010){Kim}, {Bennett}, {Chen}, {O'Brien}, \& {Kaiser}}]{Kim2010ApJ}
{Kim}, Y.~S., {Bennett}, C.~J., {Chen}, L.-H., {O'Brien}, K., \& {Kaiser}, R.~I. 2010, \apj, 711, 744

\bibitem[{{Kim} \& {Kaiser}(2009)}]{Kim2009ApJS}
{Kim}, Y.~S. \& {Kaiser}, R.~I. 2009, \apjs, 181, 543

\bibitem[{{Kobayashi} {et~al.}(2017){Kobayashi}, {Hidaka}, {Lamberts}, {Hama}, {Kawakita}, {K{\"a}stner}, \& {Watanabe}}]{Kobayashi2017ApJ}
{Kobayashi}, H., {Hidaka}, H., {Lamberts}, T., {et~al.} 2017, \apj, 837, 155

\bibitem[{{Lee} {et~al.}(2021){Lee}, {Loomis}, {Burkhardt}, {Cooke}, {Xue}, {Siebert}, {Shingledecker}, {Remijan}, {Charnley}, {McCarthy}, \& {McGuire}}]{Lee2021ApJ}
{Lee}, K. L.~K., {Loomis}, R.~A., {Burkhardt}, A.~M., {et~al.} 2021, \apjl, 908, L11

\bibitem[{{Ligterink} {et~al.}(2015){Ligterink}, {Paardekooper}, {Chuang}, {Both}, {Cruz-Diaz}, {van Helden}, \& {Linnartz}}]{Ligterink2015A&A}
{Ligterink}, N.~F.~W., {Paardekooper}, D.~M., {Chuang}, K.~J., {et~al.} 2015, \aap, 584, A56

\bibitem[{{Loomis} {et~al.}(2021){Loomis}, {Burkhardt}, {Shingledecker}, {Charnley}, {Cordiner}, {Herbst}, {Kalenskii}, {Lee}, {Willis}, {Xue}, {Remijan}, {McCarthy}, \& {McGuire}}]{Loomis2021NatAs}
{Loomis}, R.~A., {Burkhardt}, A.~M., {Shingledecker}, C.~N., {et~al.} 2021, Nature Astronomy, 5, 188

\bibitem[{{Maity} {et~al.}(2015){Maity}, {Kaiser}, \& {Jones}}]{Maity2015PCCP}
{Maity}, S., {Kaiser}, R.~I., \& {Jones}, B.~M. 2015, Physical Chemistry Chemical Physics (Incorporating Faraday Transactions), 17, 3081

\bibitem[{{Mart{\'\i}n} {et~al.}(2021){Mart{\'\i}n}, {Mangum}, {Harada}, {Costagliola}, {Sakamoto}, {Muller}, {Aladro}, {Tanaka}, {Yoshimura}, {Nakanishi}, {Herrero-Illana}, {M{\"u}hle}, {Aalto}, {Behrens}, {Colzi}, {Emig}, {Fuller}, {Garc{\'\i}a-Burillo}, {Greve}, {Henkel}, {Holdship}, {Humire}, {Hunt}, {Izumi}, {Kohno}, {K{\"o}nig}, {Meier}, {Nakajima}, {Nishimura}, {Padovani}, {Rivilla}, {Takano}, {van der Werf}, {Viti}, \& {Yan}}]{Martín2021A&A}
{Mart{\'\i}n}, S., {Mangum}, J.~G., {Harada}, N., {et~al.} 2021, \aap, 656, A46

\bibitem[{{McGuire} {et~al.}(2017){McGuire}, {Burkhardt}, {Shingledecker}, {Kalenskii}, {Herbst}, {Remijan}, \& {McCarthy}}]{McGuire2017ApJ}
{McGuire}, B.~A., {Burkhardt}, A.~M., {Shingledecker}, C.~N., {et~al.} 2017, \apjl, 843, L28

\bibitem[{{Minissale} {et~al.}(2016){Minissale}, {Moudens}, {Baouche}, {Chaabouni}, \& {Dulieu}}]{Minissale2016MNRAS}
{Minissale}, M., {Moudens}, A., {Baouche}, S., {Chaabouni}, H., \& {Dulieu}, F. 2016, \mnras, 458, 2953

\bibitem[{{Modica} \& {Palumbo}(2010)}]{Modica2010A&A}
{Modica}, P. \& {Palumbo}, M.~E. 2010, \aap, 519, A22

\bibitem[{{Molpeceres} \& {Rivilla}(2022)}]{Molpeceres2022A&A}
{Molpeceres}, G. \& {Rivilla}, V.~M. 2022, \aap, 665, A27

\bibitem[{{Nguyen} {et~al.}(2019){Nguyen}, {Fourr{\'e}}, {Favre}, {Barois}, {Congiu}, {Baouche}, {Guillemin}, {Ellinger}, \& {Dulieu}}]{Nguyen2019A&A}
{Nguyen}, T., {Fourr{\'e}}, I., {Favre}, C., {et~al.} 2019, \aap, 628, A15

\bibitem[{{Oba} {et~al.}(2018){Oba}, {Tomaru}, {Lamberts}, {Kouchi}, \& {Watanabe}}]{Oba2018NatAs}
{Oba}, Y., {Tomaru}, T., {Lamberts}, T., {Kouchi}, A., \& {Watanabe}, N. 2018, Nature Astronomy, 2, 228

\bibitem[{{{\"O}berg} {et~al.}(2011){{\"O}berg}, {Boogert}, {Pontoppidan}, {van den Broek}, {van Dishoeck}, {Bottinelli}, {Blake}, \& {Evans}}]{Oberg2011ApJ}
{{\"O}berg}, K.~I., {Boogert}, A.~C.~A., {Pontoppidan}, K.~M., {et~al.} 2011, \apj, 740, 109

\bibitem[{{{\"O}berg} {et~al.}(2009){{\"O}berg}, {Garrod}, {van Dishoeck}, \& {Linnartz}}]{Öberg2009A&A}
{{\"O}berg}, K.~I., {Garrod}, R.~T., {van Dishoeck}, E.~F., \& {Linnartz}, H. 2009, \aap, 504, 891

\bibitem[{{Pardo} {et~al.}(2005){Pardo}, {Cernicharo}, \& {Goicoechea}}]{Pardo2005ApJ}
{Pardo}, J.~R., {Cernicharo}, J., \& {Goicoechea}, J.~R. 2005, \apj, 628, 275

\bibitem[{{Pardo} {et~al.}(2022){Pardo}, {Cernicharo}, {Tercero}, {Cabezas}, {Berm{\'u}dez}, {Ag{\'u}ndez}, {Gallego}, {Tercero}, {G{\'o}mez-Garrido}, {de Vicente}, \& {L{\'o}pez-P{\'e}rez}}]{Pardo2022A&A}
{Pardo}, J.~R., {Cernicharo}, J., {Tercero}, B., {et~al.} 2022, \aap, 658, A39

\bibitem[{{Puglisi} {et~al.}(2014){Puglisi}, {Compagnini}, {D'Urso}, {Baratta}, {Palumbo}, \& {Strazzulla}}]{Puglisi2014NIMPB}
{Puglisi}, O., {Compagnini}, G., {D'Urso}, L., {et~al.} 2014, Nuclear Instruments and Methods in Physics Research B, 326, 2

\bibitem[{{Qasim} {et~al.}(2019{\natexlab{a}}){Qasim}, {Fedoseev}, {Chuang}, {Taquet}, {Lamberts}, {He}, {Ioppolo}, {van Dishoeck}, \& {Linnartz}}]{Qasim2019A&Aa}
{Qasim}, D., {Fedoseev}, G., {Chuang}, K.~J., {et~al.} 2019{\natexlab{a}}, \aap, 627, A1

\bibitem[{{Qasim} {et~al.}(2019{\natexlab{b}}){Qasim}, {Fedoseev}, {Lamberts}, {Chuang}, {He}, {Ioppolo}, {K{\"a}stner}, \& {Linnartz}}]{Qasim2019ESCb}
{Qasim}, D., {Fedoseev}, G., {Lamberts}, T., {et~al.} 2019{\natexlab{b}}, ACS Earth and Space Chemistry, 3, 986

\bibitem[{{Rubin} {et~al.}(2015){Rubin}, {Altwegg}, {Balsiger}, {Bar-Nun}, {Berthelier}, {Bieler}, {Bochsler}, {Briois}, {Calmonte}, {Combi}, {De Keyser}, {Dhooghe}, {Eberhardt}, {Fiethe}, {Fuselier}, {Gasc}, {Gombosi}, {Hansen}, {H{\"a}ssig}, {J{\"a}ckel}, {Kopp}, {Korth}, {Le Roy}, {Mall}, {Marty}, {Mousis}, {Owen}, {R{\`e}me}, {S{\'e}mon}, {Tzou}, {Waite}, \& {Wurz}}]{Rubin2015Sci}
{Rubin}, M., {Altwegg}, K., {Balsiger}, H., {et~al.} 2015, Science, 348, 232

\bibitem[{{Sakai} \& {Yamamoto}(2013)}]{Sakai2013ChRv}
{Sakai}, N. \& {Yamamoto}, S. 2013, Chemical Reviews, 113, 8981

\bibitem[{{Santos} {et~al.}(2024){Santos}, {Enrique-Romero}, {Lamberts}, {Linnartz}, \& {Chuang}}]{Santos2024ESC}
{Santos}, J.~C., {Enrique-Romero}, J., {Lamberts}, T., {Linnartz}, H., \& {Chuang}, K.-J. 2024, ACS Earth and Space Chemistry, 8, 1646

\bibitem[{{Schuhmann} {et~al.}(2019){Schuhmann}, {Altwegg}, {Balsiger}, {Berthelier}, {De Keyser}, {Fiethe}, {Fuselier}, {Gasc}, {Gombosi}, {H{\"a}nni}, {Rubin}, {Tzou}, \& {Wampfler}}]{Schuhmann2019A&A}
{Schuhmann}, M., {Altwegg}, K., {Balsiger}, H., {et~al.} 2019, \aap, 630, A31

\bibitem[{{Segr{\'e}} {et~al.}(2001){Segr{\'e}}, {Ben-Eli}, {Deamer}, \& {Lancet}}]{Segre2001OLEB}
{Segr{\'e}}, D., {Ben-Eli}, D., {Deamer}, D.~W., \& {Lancet}, D. 2001, Origins of Life and Evolution of the Biosphere, 31, 119

\bibitem[{{Stevenson} {et~al.}(2015){Stevenson}, {Lunine}, \& {Clancy}}]{Stevenson2015SciA}
{Stevenson}, J., {Lunine}, J., \& {Clancy}, P. 2015, Science Advances, 1, 1400067

\bibitem[{{Strazzulla} \& {Baratta}(1992)}]{Straz1992A&A}
{Strazzulla}, G. \& {Baratta}, G.~A. 1992, \aap, 266, 434

\bibitem[{{Taniguchi} {et~al.}(2024){Taniguchi}, {Gorai}, \& {Tan}}]{Taniguchi2024Ap&SS}
{Taniguchi}, K., {Gorai}, P., \& {Tan}, J.~C. 2024, \apss, 369, 34

\bibitem[{{Taniguchi} {et~al.}(2019){Taniguchi}, {Herbst}, {Ozeki}, \& {Saito}}]{Taniguchi2019ApJ}
{Taniguchi}, K., {Herbst}, E., {Ozeki}, H., \& {Saito}, M. 2019, \apj, 884, 167

\bibitem[{{Taniguchi} {et~al.}(2017){Taniguchi}, {Ozeki}, \& {Saito}}]{Taniguchi2017ApJ}
{Taniguchi}, K., {Ozeki}, H., \& {Saito}, M. 2017, \apj, 846, 46

\bibitem[{{Taniguchi} {et~al.}(2016){Taniguchi}, {Saito}, \& {Ozeki}}]{Taniguchi2016ApJ}
{Taniguchi}, K., {Saito}, M., \& {Ozeki}, H. 2016, \apj, 830, 106

\bibitem[{{Theule} {et~al.}(2011){Theule}, {Borget}, {Mispelaer}, {Danger}, {Duvernay}, {Guillemin}, \& {Chiavassa}}]{Theule2011A&A}
{Theule}, P., {Borget}, F., {Mispelaer}, F., {et~al.} 2011, \aap, 534, A64

\bibitem[{{T{\o}rneng} {et~al.}(1980){T{\o}rneng}, {Nielsen}, {Klaeboe}, {Hopf}, \& {Priebe}}]{Torneng1980AcSpA}
{T{\o}rneng}, E., {Nielsen}, C.~J., {Klaeboe}, P., {Hopf}, H., \& {Priebe}, H. 1980, Spectrochimica Acta Part A: Molecular Spectroscopy, 36, 975

\bibitem[{Trevors \& Psenner(2001)}]{TREVORS2001573}
Trevors, J.~T. \& Psenner, R. 2001, FEMS Microbiology Reviews, 25, 573

\bibitem[{{Tschersich}(2000)}]{Tschersich2000JAP}
{Tschersich}, K.~G. 2000, Journal of Applied Physics, 87, 2565

\bibitem[{{Turner} {et~al.}(2018){Turner}, {Abplanalp}, {Blair}, {Dayuha}, \& {Kaiser}}]{Turner2018ApJS}
{Turner}, A.~M., {Abplanalp}, M.~J., {Blair}, T.~J., {Dayuha}, R., \& {Kaiser}, R.~I. 2018, \apjs, 234, 6

\bibitem[{{Urso} {et~al.}(2019){Urso}, {Palumbo}, {Ceccarelli}, {Balucani}, {Bottinelli}, {Codella}, {Fontani}, {Leto}, {Trigilio}, {Vastel}, {Bachiller}, {Baratta}, {Buemi}, {Caux}, {Jaber Al-Edhari}, {Lefloch}, {L{\'o}pez-Sepulcre}, {Umana}, \& {Testi}}]{Urso2019A&A}
{Urso}, R.~G., {Palumbo}, M.~E., {Ceccarelli}, C., {et~al.} 2019, \aap, 628, A72

\bibitem[{{van Dishoeck} {et~al.}(2023){van Dishoeck}, {Grant}, {Tabone}, {van Gelder}, {Francis}, {Tychoniec}, {Bettoni}, {Arabhavi}, {Gasman}, {Nazari}, {Vlasblom}, {Kavanagh}, {Christiaens}, {Klaassen}, {Beuther}, {Henning}, \& {Kamp}}]{vanDishoeck2023FaDi}
{van Dishoeck}, E.~F., {Grant}, S., {Tabone}, B., {et~al.} 2023, Faraday Discussions, 245, 52

\bibitem[{{Wu} {et~al.}(2010){Wu}, {Lin}, {Chou}, {Chen}, {Lu}, {Chen}, \& {Cheng}}]{Wu2010ApJ}
{Wu}, Y.-J., {Lin}, M.-Y., {Chou}, S.-L., {et~al.} 2010, \apj, 721, 856

\bibitem[{{Yabuta} {et~al.}(2023){Yabuta}, {Cody}, {Engrand}, {Kebukawa}, {De Gregorio}, {Bonal}, {Remusat}, {Stroud}, {Quirico}, {Nittler}, {Hashiguchi}, {Komatsu}, {Okumura}, {Mathurin}, {Dartois}, {Duprat}, {Takahashi}, {Takeichi}, {Kilcoyne}, {Yamashita}, {Dazzi}, {Deniset-Besseau}, {Sandford}, {Martins}, {Tamenori}, {Ohigashi}, {Suga}, {Wakabayashi}, {Verdier-Paoletti}, {Mostefaoui}, {Montagnac}, {Barosch}, {Kamide}, {Shigenaka}, {Bejach}, {Matsumoto}, {Enokido}, {Noguchi}, {Yurimoto}, {Nakamura}, {Okazaki}, {Naraoka}, {Sakamoto}, {Connolly}, {Lauretta}, {Abe}, {Okada}, {Yada}, {Nishimura}, {Yogata}, {Nakato}, {Yoshitake}, {Iwamae}, {Furuya}, {Hatakeda}, {Miyazaki}, {Soejima}, {Hitomi}, {Kumagai}, {Usui}, {Hayashi}, {Yamamoto}, {Fukai}, {Sugita}, {Kitazato}, {Hirata}, {Honda}, {Morota}, {Tatsumi}, {Sakatani}, {Namiki}, {Matsumoto}, {Noguchi}, {Wada}, {Senshu}, {Ogawa}, {Yokota}, {Ishihara}, {Shimaki}, {Yamada}, {Honda}, {Michikami}, {Matsuoka}, {Hirata}, {Arakawa}, {Okamoto}, {Ishiguro}, {Jaumann},
  {Bibring}, {Grott}, {Schr{\"o}der}, {Otto}, {Pilorget}, {Schmitz}, {Biele}, {Ho}, {Moussi-Soffys}, {Miura}, {Noda}, {Yamada}, {Yoshihara}, {Kawahara}, {Ikeda}, {Yamamoto}, {Shirai}, {Kikuchi}, {Ogawa}, {Takeuchi}, {Ono}, {Mimasu}, {Yoshikawa}, {Takei}, {Fujii}, {Iijima}, {Nakazawa}, {Hosoda}, {Iwata}, {Hayakawa}, {Sawada}, {Yano}, {Tsukizaki}, {Ozaki}, {Terui}, {Tanaka}, {Fujimoto}, {Yoshikawa}, {Saiki}, {Tachibana}, {Watanabe}, \& {Tsuda}}]{Yabuta2023Sci}
{Yabuta}, H., {Cody}, G.~D., {Engrand}, C., {et~al.} 2023, Science, 379, abn9057

\bibitem[{{Zhou} {et~al.}(2009){Zhou}, {Kaiser}, \& {Tokunaga}}]{Zhou2009P&SS}
{Zhou}, L., {Kaiser}, R.~I., \& {Tokunaga}, A.~T. 2009, \planss, 57, 830

\bibitem[{{Zhou} {et~al.}(2010){Zhou}, {Zheng}, {Kaiser}, {Landera}, {Mebel}, {Liang}, \& {Yung}}]{Zhou2010ApJ}
{Zhou}, L., {Zheng}, W., {Kaiser}, R.~I., {et~al.} 2010, \apj, 718, 1243

\end{thebibliography}

\begin{appendix} %First online appendix
\section{QMS-TPD spectra subtraction}\label{appendixA}

\begin{figure}[h!]
\centering
\includegraphics[width=8.8cm,clip]{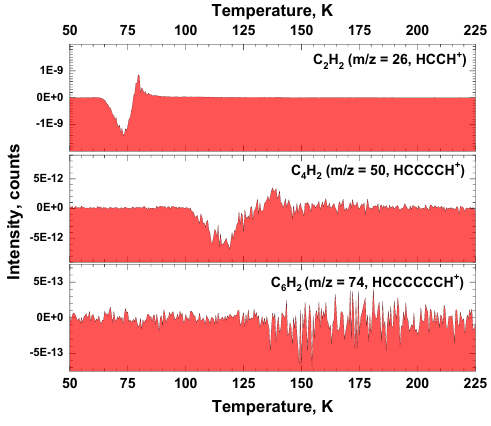}
\caption{Results of subtraction for each of the m/z values presented in Fig. \ref{fig:2}}
\label{appfigA1}
\end{figure}

\begin{figure}[h!]
\centering
\includegraphics[width=8.8cm,clip]{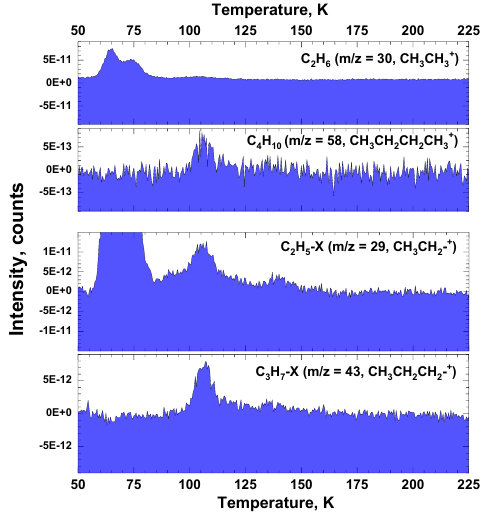}
\caption{Results of subtraction for each of the m/z values presented in Fig. \ref{fig:3}}
\label{appfigA2}
\end{figure}

\begin{figure}[h!]
\centering
\includegraphics[width=8.8cm,clip]{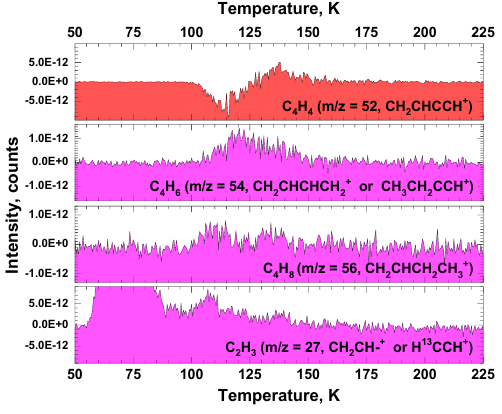}
\caption{Results of subtraction for each of the m/z values presented in Fig. \ref{fig:4}}
\label{appfigA3}
\end{figure}

Figures \ref{appfigA1}, \ref{appfigA2}, and \ref{appfigA3} show the results of subtraction between the mass spectra obtained prior and after hydrogenation of the ice for each of the individual m/z values presented through Figs. \ref{fig:2}, \ref{fig:3}, and \ref{fig:4}.

\vspace{3cm}
\section{Fragmentation pattern comparison}\label{appendixB}

\begin{figure*}[h!]
\centering
\includegraphics[width=17.6cm,clip]{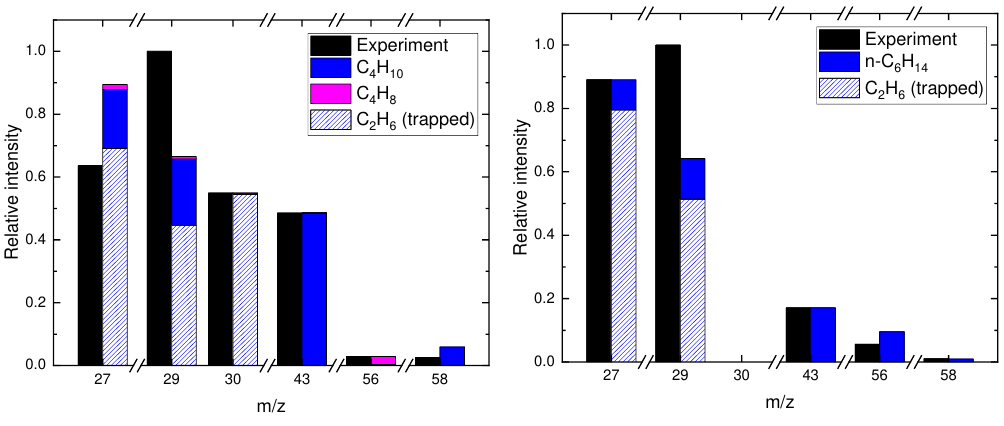}
\caption{Left: Comparison between the observed intensities of m/z signals at 105 K for 70 eV electron ionization energy of the QMS and the mass fragmentation patterns of C$_4$H$_{10}$ (CH$_3$CH$_2$CH$_2$CH$_3$), C$_4$H$_8$ (CH$_2$CHCH$_2$CH$_3$), and C$_2$H$_6$ (CH$_3$CH$_3$) acquired from NIST database. Right: Similar comparison between intensities obtained at 138 K with the mass fragmentation patterns of C$_6$H$_{14}$ (CH$_3$CH$_2$CH$_2$CH$_2$CH$_2$CH$_3$) and C$_2$H$_6$ (CH$_3$CH$_3$) acquired from the NIST database. Experimental values are baseline subtracted. Only the selected m/z values with resolved peaks are presented.}
\label{appfigB}
\end{figure*}

In Fig. \ref{appfigB}, examples of comparison between the m/z signals registered at 105 and 138 K with the mass spectra available from the NIST database are presented. It is shown that the ratio between m/z$\geq$43 signals demonstrates relatively good agreement with the NIST spectra of linear n-C$_4$H$_{10}$ and n-C$_6$H$_{14}$. However, co-desorption of entrapped C$_2$H$_6$ (or other species) along with the main constituents of the ice is required to account for the high relative intensity of m/z= 27, 29, 30 signals registered at these temperatures.

\end{appendix}
\end{document}